\documentclass[aps,pra,twocolumn,longbibliography]{revtex4-1}

\usepackage{hyperref}
\hypersetup{
  breaklinks=true,
  colorlinks=true,
  linkcolor=blue,
  urlcolor=cyan,
}

\usepackage{physics,braket,bm,commath,amssymb}
\renewcommand{\t}{\text} % text in math mode
\newcommand{\f}[2]{\dfrac{#1}{#2}} % shorthand for fractions
\newcommand{\p}[1]{\left( #1 \right)} % parenthesis
\renewcommand{\sp}[1]{\left[ #1 \right]} % square parenthesis
\renewcommand{\v}{\bm} % bold vectors
\newcommand{\uv}[1]{\bm{\hat{#1}}} % unit vectors

\renewcommand{\abs}[1]{\lvert #1 \rvert}

\newcommand{\bk}{\braket} % shorthand for braket notation
\newcommand{\Bk}{\Braket}

\newcommand{\ul}{\underline} % shorthand for underline

\newcommand{\D}{\mathcal{D}}
\newcommand{\E}{\mathcal{E}}
\newcommand{\J}{\mathcal{J}}
\renewcommand{\O}{\mathcal{O}}
\newcommand{\Q}{\mathcal{Q}}
\renewcommand{\S}{\mathcal{S}}
\newcommand{\U}{\mathcal{U}}

\newcommand{\C}{\mathbb{C}}
\newcommand{\N}{\mathbb{N}}

\newcommand{\z}{\text{z}}
\newcommand{\x}{\text{x}}
\newcommand{\y}{\text{y}}
\newcommand{\Z}{\text{Z}}
\newcommand{\X}{\text{X}}

\newcommand{\bmu}{{\bar\mu}}
\newcommand{\bnu}{{\bar\nu}}

\usepackage{dsfont} % for identity operator
\newcommand{\1}{\mathds{1}}

\newcommand{\up}{\uparrow}
\newcommand{\dn}{\downarrow}

\renewcommand{\a}{\alpha} % free index
\renewcommand{\b}{\beta} % free index

\usepackage{graphicx} % for figures
\usepackage[caption=false]{subfig} % subfigures (via \subfloat[]{})
\newcommand{\sref}[1]{\protect\subref{#1}} % for referencing subfigures
\usepackage[inline]{enumitem} % for inline enumeration
\setenumerate{label=(\roman*)}

\begin{document}

\title{Short-time expansion of Heisenberg operators in open collective
  quantum spin systems}

\author{Michael A. Perlin}
\email{mika.perlin@gmail.com}
\author{Ana Maria Rey}
\affiliation{JILA, National Institute of Standards and Technology and
  University of Colorado, 440 UCB, Boulder, Colorado 80309, USA}
\affiliation{Center for Theory of Quantum Matter, 440 UCB, Boulder,
  Colorado 80309, USA}
\affiliation{Department of Physics, University of Colorado, 390 UCB,
  Boulder, Colorado 80309, USA}

\begin{abstract}
  We present an efficient method to compute short-time expectation
  values in large collective spin systems with typical forms of
  Markovian decoherence.  Our method is based on a Taylor expansion of
  a formal solution to the equations of motion for Heisenberg
  operators.  This expansion can be truncated at finite order to
  obtain virtually exact results at short times that are relevant for
  metrological applications such as spin squeezing.  In order to
  evaluate the expansion for Heisenberg operators, we compute the
  relevant structure constants of a collective spin operator algebra.
  We demonstrate the utility of our method by computing spin
  squeezing, two-time correlation functions, and out-of-time-ordered
  correlators for $10^4$ spins in strong-decoherence regimes that are
  otherwise inaccessible via existing numerical methods.  Our method
  can be straightforwardly generalized to the case of a collective
  spin coupled to bosonic modes, relevant for trapped ion and cavity
  QED experiments, and may be used to investigate short-time
  signatures of quantum chaos and information scrambling.
\end{abstract}

\maketitle

%%%%%%%%%%%%%%%%%%%%%%%%%%%%%%%%%%%%%%%%%%%%%%%%%%%%%%%%%%%%%%%%%%%%%%
\section{Introduction}

Collective spin systems are a versatile resource in quantum science
for a range of applications including quantum-enhanced metrology and
quantum simulation.  The study of such systems dates back to the
mid-twentieth century with the introduction of the Dicke
model\cite{dicke1954coherence} that describes atoms cooperatively
interacting with a single mode of a radiation field, and the
Lipkin-Meshkov-Glick (LMG) model, a toy model for testing many-body
approximation methods in contemporary nuclear
physics\cite{lipkin1965validity, meshkov1965validity,
  glick1965validity}.  On the experimental side, the development of
advanced trapping, cooling, and control techniques in atomic,
molecular, and optical (AMO) systems have enabled the realization of
collective spin models in a broad range of platforms, including cold
atomic gasses\cite{takano2009spin, appel2009mesoscopic}, Bose-Einstein
condensates\cite{klinder2015dynamical, esteve2008squeezing,
  riedel2010atomchipbased, gross2010nonlinear}, ultracold Fermi
gasses\cite{martin2013quantum, bromley2018dynamics,
  smale2019observation}, trapped ions\cite{bohnet2016quantum}, and
optical cavities\cite{schleier-smith2010states, chen2011conditional,
  baumann2010dicke, leroux2010implementation, bohnet2014reduced,
  cox2016deterministic, hosten2016measurement, hosten2016quantum,
  norcia2018cavitymediated, ritsch2013cold}, among others.  These
implementations compliment innumerable theoretical studies in a
variety of rich subjects, including quantum phase transitions and
criticality\cite{latorre2005entanglement, alcalde2007functional,
  wang2012quantum, majd2014lmg}, non-equilibrium
phenomena\cite{walls1978nonequilibrium, morrison2008dynamical,
  morrison2008dissipationdriven, morrison2008collective,
  kessler2012dissipative, bhattacherjee2014nonequilibrium,
  zhiqiang2017nonequilibrium, lang2018concurrence}, and precision
mentrology\cite{wineland1992spin, kitagawa1993squeezed,
  zhong2010simplified, schleier-smith2010squeezing, ma2011quantum,
  huang2015twoaxis, muessel2015twistandturn, huang2015quantum,
  hu2017vacuum, mirkhalaf2018robustifying, lewis-swan2018robust,
  he2019engineering}.

One of the primary motivations for studying collective spin systems is
their application to quantum-enhanced metrology.  Quantum projection
noise limits the error $\Delta\phi$ in the measurement of a phase
angle $\phi$ with $N$ independent spins to
$\Delta\phi\sim1/\sqrt{N}$\cite{wineland1992spin, itano1993quantum}.
Collective spin systems provide a means to break through this limit
via the preparation of many-body entangled states such as spin-cat
states\cite{agarwal1997atomic, lau2014proposal, huang2015quantum} and
most notably spin-squeezed states\cite{wineland1992spin,
  kitagawa1993squeezed, ma2011quantum} that allow for measurement
errors $\Delta\phi\sim1/N^\varepsilon$ with $1/2<\varepsilon\le1$,
where $\varepsilon=1$ saturates the Heisenberg
limit\cite{zwierz2010general}.  Such entangled states can be prepared
either via heralded methods such as quantum nondemolition
measurements\cite{takano2009spin, appel2009mesoscopic,
  schleier-smith2010states, chen2011conditional}, or via deterministic
methods that require nonlinear dynamics, typically realized with
phonon-mediated\cite{bohnet2016quantum},
photon-mediated\cite{klinder2015dynamical, baumann2010dicke,
  leroux2010implementation, bohnet2014reduced, cox2016deterministic,
  hosten2016measurement, hosten2016quantum, norcia2018cavitymediated,
  ritsch2013cold} or collisional\cite{esteve2008squeezing,
  riedel2010atomchipbased, gross2010nonlinear, martin2013quantum,
  bromley2018dynamics, smale2019observation} interactions.  Although a
truly collective spin model requires uniform, all-to-all interactions,
as long as measurements do not distinguish between constituent
particles, even non-uniform systems may be effectively described by a
uniform model with renormalized parameters\cite{hu2015entangled}.

In the absence of decoherence, permutation symmetry and total spin
conservation divide the total Hilbert space of a collective spin
system into superselection sectors that grow only linearly with system
size $N$, thereby admitting efficient classical simulation of its
dynamics.  Decoherence generally violates total spin conservation and
requires the use of density operators, increasing the dimension of
accessible state space to $O\p{N^3}$\cite{hartmann2016generalized,
  xu2013simulating}.  In this case, exact simulations can be carried
out for $N\lesssim100$ particles.  If decoherence is sufficiently
weak, dynamics can be numerically solvable for $N\lesssim10^5$
particles via ``quantum trajectory'' Monte Carlo
methods\cite{plenio1998quantumjump, zhang2018montecarlo} (also known
as ``quantum jump'' or ``Monte Carlo wavefunction'' methods) that can
reproduce all expectation values of interest.  When decoherence is
strong, however, these Monte Carlo methods can take a prohibitively
long time to converge, as simulations become dominated by incoherent
jumps that generate large numbers of distinct quantum trajectories
that need to be averaged in order to accurately compute expectation
values.  Even with strong decoherence, dynamics are sometimes solvable
through the cumulant expansion\cite{meiser2010steadystate} that
neglects all $n$-body connected correlators for $n>2$.  The growth of
genuinely multi-body correlations, however, eventually causes the
cumulant expansion to yield incorrect results with no clear signature
of failure.  In the absence of other means to compute correlators, it
can therefore be difficult to identify the point at which correlators
computed via the cumulant expansion can no longer be trusted.

In this work, we present an efficient method to compute short-time
dynamics of collective spin systems with typical forms of Markovian
decoherence.  The only restriction on decoherence (beyond
Markovianity) is that, like the coherent collective dynamics, it must
act identically on all constituent particles.  Our method is based on
a formal solution to the equations of motion for Heisenberg operators,
thereby bearing some resemblance to the Mori
formalism\cite{mori1965continuedfraction} and related
work\cite{annett1994recursive}.  Specifically, we expand a formal
solution for a Heisenberg operator into a Taylor series whose
truncation can yield negligible error at sufficiently short times.
Evaluating the resulting expansion requires knowing the structure
constants of a collective spin operator algebra; the calculation of
these structure constants (in Appendices
\ref{sec:identities}--\ref{sec:general_product}) is one of the main
technical results of this work, which we hope will empower both
analytical and numerical studies of collective spin systems in the
future.  We benchmark our method against exact results from both
analytical calculations and quantum trajectory Monte Carlo
computations of spin squeezing in accessible parameter regimes,
highlighting both advantages and limitations of the short-time
expansion.  Finally, we showcase applications of our method by
computing quantities that are inaccessible to other numerical methods.

%%%%%%%%%%%%%%%%%%%%%%%%%%%%%%%%%%%%%%%%%%%%%%%%%%%%%%%%%%%%%%%%%%%%%%
\section{Theory}
\label{sec:theory}

In this section we provide the basic theory for our method to compute
expectation values of collective spin operators, deferring lengthy
derivations to the appendices.  We consider a system of $N$ distinct
spin-1/2 particles.  Defining individual spin-1/2 operators
$\hat s_{\a=\x,\y,\z}\equiv\hat\sigma_\a/2$ and
$\hat s_\pm\equiv\hat s_\x\pm i\hat s_\y=\hat\sigma_\pm$ with Pauli
operators $\hat\sigma_\a$, we denote an operator that acts with
$\hat s_\a$ on the spin indexed by $j$ and trivially (i.e.~with the
identity $\hat\1$) on all other spins by $\hat s_\a^{(j)}$.  We then
define the collective spin operators
$\hat S_\a\equiv\sum_{j=1}^N\hat s_\a^{(j)}$ for
$\a\in\set{\x,\y,\z,+,-}$.  Identifying the set $\set{\hat\S_{\v m}}$
as a basis for all collective spin operators, with
$\v m\equiv\p{m_+,m_\z,m_-}\in\N_0^3$ and
$\hat\S_{\v m}\equiv \hat S_+^{m_+} \hat S_\z^{m_\z} \hat S_-^{m_-}$,
we can expand any collective spin operator $\hat\O$ in the form
\begin{align}
  \hat\O = \sum_{\v m} \O_{\v m} \hat\S_{\v m}
\end{align}
with scalar coefficients $\O_{\v m}\in\C$.  If $\hat\O$ is
self-adjoint, for example, then $\O_{\v m}^*=\O_{\v m^*}$ with
$\v m^*\equiv\p{m_-,m_\z,m_+}$.  The corresponding Heisenberg operator
is then
$\hat\O\p{t}=\sum_{\v m}\O_{\v m}\p{t}\hat\S_{\v m}+\hat\E_\O\p{t}$,
with time-dependent coefficients $\O_{\v m}\p{t}$ for time-independent
Schr\"odinger operators $\hat\S_{\v m}$, and mean-zero ``noise''
operators $\hat\E_\O\p{t}$ that result from interactions between the
spin system and its environment, initially $\hat\E_\O\p{0}=0$.  These
noise operators will essentially play no role in the present work, but
are necessary to include for a consistent formalism of Heisenberg
operators in an open quantum system; see Appendix \ref{sec:noise} for
further discussion.  The expectation values of Heisenberg operators
evolve according to
\begin{align}
  \f{d}{dt} \bk{\hat\O\p{t}}
  = \bk{\check T \hat\O\p{t}}
  = \sum_{\v m, \v n} \bk{\hat\S_{\v m}} T_{\v m\v n} \O_{\v n}\p{t}
  \label{eq:EOM}
\end{align}
with a Heisenberg-picture time derivative operator $\check T=d/dt$
whose matrix elements $T_{\v m\v n}\in\C$ are defined by
\begin{align}
  \check T \hat\S_{\v n} \equiv i \sp{\hat H, \hat\S_{\v n}}_-
  + \sum_\J \gamma_\J \check\D\p{\J} \hat\S_{\v n}
  = \sum_{\v m} \hat\S_{\v m} T_{\v m\v n},
  \label{eq:time_deriv}
\end{align}
where $\sp{X,Y}_\pm\equiv XY\pm YX$; $\hat H$ is the collective spin
Hamiltonian; $\J$ is a set of jump operators with a corresponding
decoherence rate $\gamma_\J$; and $\check\D$ is a Heisenberg-picture
dissipator, or Lindblad superoperator, defined by
\begin{align}
  \check\D\p{\J} \hat \O
  \equiv \sum_{\hat J\in\J} \p{\hat J^\dag \hat \O \hat J
    - \f12\sp{\hat J^\dag \hat J, \hat \O}_+}.
\end{align}
Decoherence via uncorrelated decay of individual spins, for example,
would be described by the set of jump operators
$\J_-\equiv\set{\hat s_-^{(j)}:j=1,2,\cdots,N}$.  The commutator in
Eq.~\eqref{eq:time_deriv} can be computed by expanding the product
$\hat\S_{\v\ell}\hat\S_{\v m} = \sum_{\v n} f_{\v\ell\v m\v
  n}\hat\S_{\v n}$ with structure constants
$f_{\v\ell\v m\v n}\in\mathbb{R}$ that we work out in Appendices
\ref{sec:identities}--\ref{sec:general_product}, and the effects of
decoherence from jump operators (i.e.~elements of $\J$) of the form
$\hat g^{(j)} = \sum_\a g_\a \hat s_\a^{(j)}$ and
$\hat G = \sum_\a G_\a \hat S_\a$ are worked out in Appendices
\ref{sec:sandwich_single}--\ref{sec:decoherence_collective}.  We
consider these calculations to be some of the main technical
contributions of this work, with potential applications beyond the
short-time simulation method presented here.  These ingredients are
sufficient to compute matrix elements $T_{\v m\v n}$ of the time
derivative operator $\check T$ in Eq.~\eqref{eq:time_deriv} in most
cases of practical interest.

We note that particle loss is an important decoherence mechanism in
many experimental realizations of collective spin
models\cite{ma2011quantum}.  In principle, a spin model has no notion
of the particle annihilation operators that generate particle loss,
and therefore cannot capture this effect directly.  Nonetheless, for a
system initially composed of $N$ particles, the effect of particle
loss can be emulated with $O(1/N)$ error by the dissipator
$\check\D_{\t{loss}}$ defined by
$\check\D_{\t{loss}}\hat\S_{\v m}=-\abs{\v m}\hat\S_{\v m}$, where
$\abs{\v m}\equiv\sum_\alpha m_\alpha$ (see Appendix
\ref{sec:particle_loss}).  Furthermore, the effect particle loss can
be accounted for exactly by
\begin{enumerate*}
\item introducing an additional index on spin operators,
  $\hat\S_{\v m}\to\hat\S_{N\v m}$, to keep track of different sectors
  of fixed particle number within a multi-particle Fock space, and
\item constructing jump operators that appropriately couple spin
  operators within different particle-number sectors.
\end{enumerate*}
We defer a detailed exact accounting of particle loss to future work.

The time derivative operator $\check T$ will generally couple spin
operators $\hat\S_{\v n}$ to spin operators $\hat\S_{\v m}$ with
higher ``weight'', i.e.~with $\abs{\v m}>\abs{\v n}$.  The growth of
operator weight signifies the growth of many-body correlations.
Keeping track of this growth eventually becomes intractable, requiring
us to truncate our equations of motion somehow.  The simplest
truncation strategy would be to take
\begin{align}
  \f{d}{dt} \bk{\hat\O\p{t}}
  \to \sum_{w\p{\v m}<W} \bk{\hat\S_{\v m}}
  \sum_{\v n} T_{\v m\v n} \O_{\v n}\p{t}
  \label{eq:weight_truncation}
\end{align}
for some weight measure $w$, e.g.~$w\p{\v m}=\abs{\v m}$, and a
high-weight cutoff $W$.  The truncation in
Eq.~\eqref{eq:weight_truncation} closes the system of differential
equations defined by Eq.~\eqref{eq:EOM}, and allows us to solve it
using standard numerical methods.  Some initial conditions for this
system of differential equations, namely expectation values of
collective spin operators with respect to spin-polarized (Gaussian)
states that are generally simple to prepare experimentally, are
provided in Appendix \ref{sec:initial_conditions}.

The truncation strategy in Eq.~\eqref{eq:weight_truncation} has a few
limitations:
\begin{enumerate*}
\item simulating a system of differential equations for a large number
  of operators can be time-consuming,
\item the weight measure $w$ may need to be chosen carefully, as the
  optimal measure is generally system-dependent, and
\item simulation results can only be trusted up to the time at which
  the initial values of operators $\hat\S_{\v m}$ with weight
  $w\p{\v m}\ge W$ have a non-negligible contribution to expectation
  values of interest.\label{pt:limitation}
\end{enumerate*}
The last limitation in particular unavoidably applies in some form to
any method tracking only a subset of all relevant operators.  We
therefore devise an alternate truncation strategy built around
limitation \ref*{pt:limitation}.

We can formally expand Heisenberg operators $\hat\O\p{t}$ in a Taylor
series about the time $t=0$ to write
\begin{align}
  \bk{\hat\O\p{t}}
  = \bk{e^{t\check T} \hat \O\p{0}}
  = \sum_{k\ge0} \f{t^k}{k!}
  \sum_{\v m, \v n} \bk{\hat\S_{\v m}} T^k_{\v m\v n} \O_{\v n}\p{0},
  \label{eq:time_series}
\end{align}
where the matrix elements $T^k_{\v m\v n}$ of the $k$-th time
derivative operator $\check T^k$ are
\begin{align}
  T^0_{\v m\v n} &\equiv \delta_{\v m\v n}, \\
  T^1_{\v m\v n} &\equiv T_{\v m\v n}, \\
  T^{k>1}_{\v m\v n}
  &\equiv \sum_{\v p_1,\v p_2,\cdots,\v p_{k-1}}
  T_{\v m\v p_{k-1}} \cdots T_{\v p_3\v p_2}
  T_{\v p_2\v p_1} T_{\v p_1\v n},
\end{align}
with $\delta_{\v m\v n}=1$ if $\v m=\v n$ and zero otherwise.  For
sufficiently short times, we can truncate the series in
Eq.~\eqref{eq:time_series} by taking
\begin{align}
  \bk{\hat\O\p{t}}
  \to \sum_{k=0}^M \f{t^k}{k!}
  \sum_{\v m, \v n} \bk{\hat\S_{\v m}} T^k_{\v m\v n} \O_{\v n}\p{0}.
  \label{eq:TST}
\end{align}
We refer to Eq.~\eqref{eq:TST} as the truncated short-time (TST)
expansion of Heisenberg operators.  Note that when computing an
expectation value $\bk{\hat\O\p{t}}$, the relation
$\hat\S_{\v m}^\dag=\hat\S_{\v m^*}$, which by Hermitian conjugation
of Eq.~\eqref{eq:EOM} also implies that
$T_{\v m^*\v n^*}=T_{\v m\v n}^*$, cuts both the number of
initial-time expectation values $\bk{\hat\S_{\v m}}$ and the number of
matrix elements $T_{\v m\v n}$ that we may need to explicitly compute
roughly in half.

Unlike the weight-based truncation in
Eq.~\eqref{eq:weight_truncation}, the nonzero matrix elements
$T^k_{\v m\v n}$ for $k=0,1,\cdots,M$ in Eq.~\eqref{eq:TST} tell us
which operators $\hat\S_{\v m}$ are relevant for computing the
expectation value $\bk{\hat\O\p{t}}$ to a fixed order $M$.  The TST
expansion thereby avoids the introduction of a weight measure $w$ that
chooses which operators to keep track of, and trades the cost of
solving a system of differential equations for the cost of computing
expectation values $\bk{\hat\S_{\v m}}$ and matrix elements
$T_{\v m\v n}^k$.  In all cases considered in this work, we find that
the TST expansion is both faster to evaluate and provides accurate
correlators $\bk{\hat\O\p{t}}$ until later times $t$ than the
weight-based expansion in \eqref{eq:weight_truncation} with weight
measure $w\p{\v m}=\abs{\v m}$ and cutoff $W\approx M$.  We therefore
restrict the remainder of our discussions to the TST expansion in
Eq.~\eqref{eq:TST}, and provide a pedagogical tutorial for computing
correlators using the TST expansion in Appendix \ref{sec:tutorial}.

Three primary considerations limit the maximum time $t$ to which we
can accurately compute a correlator $\bk{\hat S_{\v n}\p{t}}$ using
the TST expansion.  First, maintaining accuracy at larger times $t$
requires going to higher orders $M$ in the TST expansion.  An
order-$M$ TST expansion of the correlator $\bk{\hat S_{\v n}\p{t}}$
can involve a significant fraction of operators $\hat S_{\v m}$ with
weight $\abs{\v m}\lesssim M$, which implies the need to compute
$O\p{M^3}$ initial-time expectation values $\bk{\hat S_{\v m}}$ and
$O\p{M^4}$ matrix elements $T^k_{\v m\v n}$.  In practice, with a
straightforward implementation of the TST expansion we find that these
requirements generally restrict $M\lesssim 50$ -- $70$ with $8$ --
$50$ gigabytes of random access memory (RAM).  Second, individual
terms at high orders of the TST expansion in Eq.~\eqref{eq:TST} can
grow excessively large, greatly amplifying any numerical errors and
thereby spoiling cancellations that are necessary to arrive at a
physical value of a correlator, i.e.~with
$\abs{\bk{\hat S_{\v n}\p{t}}}\lesssim S^{\abs{\v n}}$ (where
$S\equiv N/2$).  Finally, the TST expansion is essentially
perturbative in the time $t$, which implies that its validity as a
formal expansion eventually breaks down.  Precisely characterizing the
implications of these last two considerations for the TST expansion
requires additional analysis that we defer to future work.  An
investigation of connections between the TST expansion and past work
related to the Mori formalism\cite{mori1965continuedfraction,
  annett1994recursive}, for example, might answer questions about the
breakdown and convergence of the TST expansion.  As we show from
benchmarks of the TST expansion in Section \ref{sec:squeezing},
however, a detailed understanding of breakdown is not necessary to
diagnose the breakdown time $t_{\t{break}}^{(M)}$ beyond which the TST
expansion yields inaccurate results.  Empirically, we find that going
beyond order $M\approx35$ yields no significant gains in all cases
considered in this work.

%%%%%%%%%%%%%%%%%%%%%%%%%%%%%%%%%%%%%%%%%%%%%%%%%%%%%%%%%%%%%%%%%%%%%%
\section{Spin squeezing, benchmarking, and breakdown}
\label{sec:squeezing}

To benchmark our method for computing collective spin correlators, we
consider three collective spin models known to generate spin-squeezed
states: the one-axis twisting (OAT), two-axis twisting (TAT), and
twist-and-turn (TNT) models described by the collective spin
Hamiltonians\cite{ma2011quantum}
\begin{align}
  H_{\t{OAT}} &= \chi \hat S_\z^2, \label{eq:OAT} \\
  H_{\t{TAT}}
  &= \f{\chi}{3} \p{\hat S_\z^2 - \hat S_\y^2}, \label{eq:TAT} \\
  H_{\t{TNT}} &= \chi \hat S_\z^2 + \Omega \hat S_\x, \label{eq:TNT}
\end{align}
where we include a factor of $1/3$ in the TAT Hamiltonian because it
naturally appears in realistic proposals to experimentally implement
TAT\cite{liu2011spin, huang2015twoaxis}.  For simplicity, we further
fix $\Omega=\chi S$ (with $S\equiv N/2$ throughout this work) to the
critical value known to maximize the entanglement generation rate of
TNT in the large-$N$ limit\cite{micheli2003manyparticle,
  sorelli2019fast}.

Note that the OAT model is a special case of the zero-field Ising
model, whose quantum dynamics admits an exact analytic solution even
in the presence of decoherence\cite{foss-feig2013nonequilibrium}.  The
approximate and numerics-oriented TST expansion is therefore an
inappropriate tool for studying the OAT model, which will merely serve
as an exactly solvable benchmark of our methods.  Wherever applicable,
we will provide exact results for the OAT model (see Appendix
\ref{sec:OAT}, as well as the Supplementary Material of
Ref.~[\citenum{bohnet2016quantum}]).

The Hamiltonians in Eqs.~\eqref{eq:OAT}--\eqref{eq:TNT} squeeze the
initial product state $\ket\X\propto\p{\ket\up+\ket\dn}^{\otimes N}$
with $\hat S_\x\ket\X=S\ket\X$.  Our measure of spin squeezing is the
directionally-unbiased Ramsey squeezing parameter determined by the
maximal gain in resolution $\Delta\phi$ of a phase angle $\phi$ over
that achieved by any spin-polarized product state
(e.g.~$\ket\X$)\cite{wineland1992spin, ma2011quantum},
\begin{align}
  \xi^2
  \equiv \f{\p{\Delta\phi_{\t{min}}}^2}{\p{\Delta\phi_{\t{polarized}}}^2}
  = \f{N}{\abs{\braket{\uv S}}^2}
  \min_{\substack{\v v\perp\bk{\uv S}\\\v v\cdot\v v=1}}
  \Bk{\p{\uv S\cdot\v v}^2},
  \label{eq:squeezing}
\end{align}
where $\uv S\equiv\p{\hat S_\x,\hat S_\y,\hat S_\z}$ is a collective
spin operator-valued vector, the minimization is performed over all
unit vectors $\v v$ orthogonal to the mean spin vector $\bk{\uv S}$,
and for brevity we have suppressed the explicit time dependence of
operators in Eq.~\eqref{eq:squeezing}.  This squeezing parameter is
entirely determined by one- and two-spin correlators of the form
$\bk{\hat S_\a}$ and $\bk{\hat S_\a \hat S_\b}$.  For the unitary
dynamics discussed in this work, these correlators are obtainable via
exact simulations of quantum dynamics in the $\p{N+1}$-dimensional
Dicke manifold of states $\set{\ket{S,m}}$ with net spin $S$ and spin
projection $m$ onto the $z$ axis, i.e.~with
$\bk{S,m|\uv S^2|S,m}=S\p{S+1}$ and $\bk{S,m|\hat S_\z|S,m}=m$ for
$m\in\set{-S,-S+1,\cdots,S}$.  In the presence of single-spin or
collective decoherence, meanwhile, these correlators are obtainable
with the collective-spin quantum trajectory Monte Carlo method
developed in ref.~\cite{zhang2018montecarlo}.  In this work, these
exact and quantum trajectory simulations will be used to benchmark the
TST expansion in Eq.~\eqref{eq:TST}.

\begin{figure}
  \centering
  \subfloat[Squeezing with unitary dynamics
  \label{fig:squeezing_unitary}]
  {\includegraphics{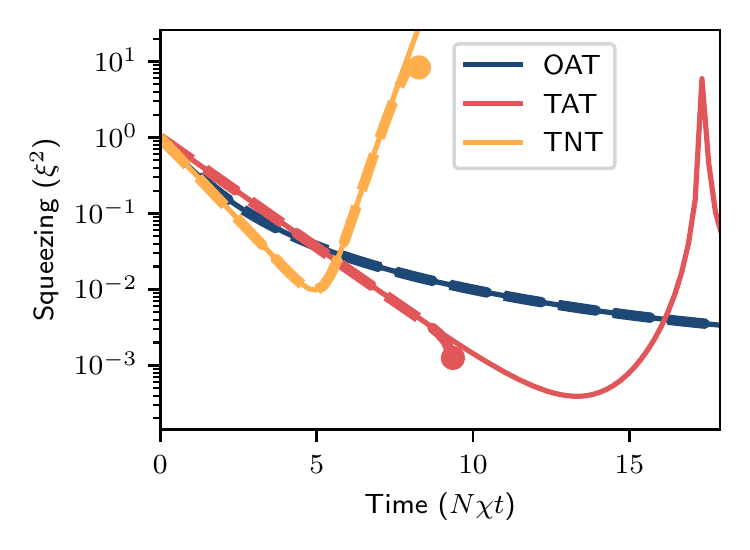}} \\
  \subfloat[Squeezing with decoherence:
  $\gamma_-=\gamma_+=\gamma_\z=\chi$
  \label{fig:squeezing_dec}]
  {\includegraphics{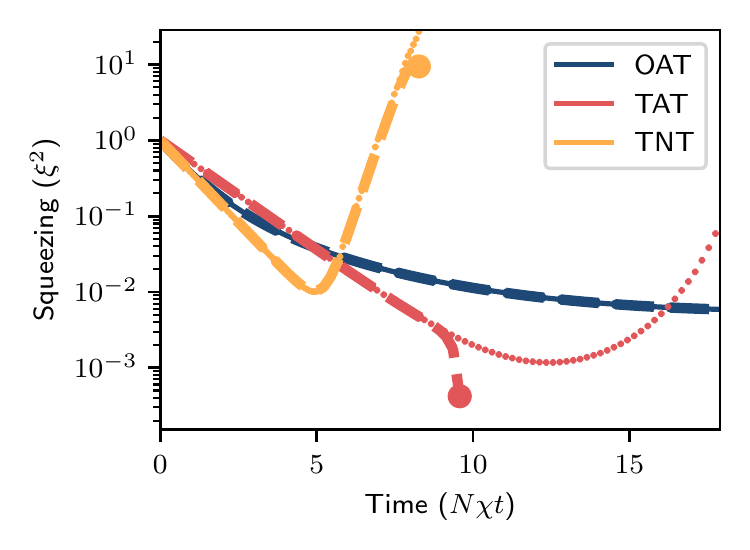}}
  \caption{Spin squeezing of $N=10^4$ spins initially in $\ket\X$
    under \sref{fig:squeezing_unitary} unitary and
    \sref{fig:squeezing_dec} non-unitary dynamics, computed using
    exact methods (solid lines), quantum trajectory simulations
    (dots), and the TST expansion in Eq.~\eqref{eq:TST} with $M=35$
    (dashed lines).  Solid circles mark the times at which the TST
    expansion gives an unphysical result with $\xi^2<0$.}
  \label{fig:benchmarking}
\end{figure}

Figure \ref{fig:benchmarking} compares the squeezing parameter $\xi^2$
for $N=10^4$ spins initially in the state $\ket\X$ evolved under the
Hamiltonians in Eqs.~\eqref{eq:OAT}--\eqref{eq:TNT}, as computed via
both benchmarking simulations and the TST expansion in
Eq.~\eqref{eq:TST} with $M=35$.  Squeezing is shown for both unitary
dynamics (Figure \ref{fig:squeezing_unitary}), as well as non-unitary
dynamics in the presence of spontaneous decay, excitation, and
dephasing of individual spins at rates $\chi$ (Figure
\ref{fig:squeezing_dec}), respectively described by the sets of jump
operators $\J_\a\equiv\set{\hat s_\a^{(j)}}$ with corresponding
decoherence rates $\gamma_\a=\chi$ for $\a\in\set{-,+,\z}$.  The
results shown in Figure \ref{fig:benchmarking} were computed in a
rotated basis with
$\p{\hat s_\z,\hat s_\x}\to\p{\hat s_\x,-\hat s_\z}$ and
$\ket\X\to\ket{-\Z}\equiv\ket\dn^{\otimes N}$, as well as appropriate
transformations of the Hamiltonian and jump operators.  The only
effects of this rotation on the results presented in Figure
\ref{fig:benchmarking} are to
\begin{enumerate*}
\item reduce the time it takes to compute correlators
  $\bk{\hat\O\p{t}}$ with the TST expansion, and
\item prolong the time for which the TST expansion of TNT results
  agree with benchmarking simulations.
\end{enumerate*}
The speedup in a different basis occurs because for the initial state
$\ket{-\Z}$, all initial-time correlators $\bk{\hat S_{\v m}}$ are
zero unless $m_+=m_-=0$, and all non-zero correlators take $O\p{1}$
(i.e.~constant in $N$) time to compute, rather than $O\p{N}$ time (see
Appendix \ref{sec:initial_conditions}).  In total, the use of a
rotated basis reduces the computation time of initial-time correlators
from $O\p{M^3N}$ to $O\p{M}$.  The reason for prolonged agreement of
TNT results in a rotated basis is not entirely understood, and
provides a clue into the precise mechanism by which the TST expansion
breaks down (discussed below).  We defer a detailed study of this
breakdown to future work.

The main lesson from Figure \ref{fig:benchmarking} is that the TST
expansion yields essentially exact results right up until a sudden and
drastic departure that can be diagnosed by inspection.  The breakdown
of the TST expansion in Figure \ref{fig:benchmarking} induces an
unphysical squeezing parameter $\xi^2<0$.  In general, however, there
is no fundamental relationship between the breakdown of the TST
expansion and the conditions for a physical squeezing parameter
$\xi^2$.  A proper diagnosis of breakdown therefore requires
inspection of the correlators $\bk{\hat\S_{\v n}\p{t}}$ used to
compute the squeezing parameter $\xi^2$, which upon breakdown will
rapidly take unphysical values with
$\abs{\bk{\hat\S_{\v n}\p{t}}}\gtrsim S^{\abs{\v n}}$ (see Appendix
\ref{sec:breakdown} for an example).  The sudden and drastic departure
from virtually exact results is consistent with the limitations of the
TST expansion discussed at the end of Section \ref{sec:theory}.
Specifically, we identify three possible mechanisms for breakdown:
\begin{enumerate*}
\item a rapid growth in the order $M$ necessary for the TST expansion
  to converge,
\item the growth of numerical errors in excessively large terms of the
  TST expansion, and
\item the formal breakdown of the perturbative expansion in the time
  $t$.
\end{enumerate*}
In all of these cases, a detailed cancellation eventually ceases to
occur between large terms at high orders in the TST expansion.  These
large terms grow with the time $t$ raised to some large power (as high
as $M$), and therefore rapidly yield wildly unphysical results.  In
contrast to other approximate methods such as the cumulant
expansion\cite{meiser2010steadystate}, the TST expansion can thus
diagnose its own breakdown, which is an important feature when working
in parameter regimes that are inaccessible via other means to compute
correlators.  Note that, due to the breakdown mechanisms of the TST
expansion, going up through order $M=70$ does not significantly
increase the breakdown time $t_{\t{break}}^{(M)}$ in Figure
\ref{fig:benchmarking}, and in some cases even shortens
$t_{\t{break}}^{(M)}$.

Although the TST expansion breaks down at short times, it has two key
advantages over the quantum trajectory Monte Carlo method to compute
correlators in the presence of decoherence.  First, computing spin
correlators with the TST expansion is generally faster and requires
fewer computing resources.  The TST expansion results in Figure
\ref{fig:squeezing_dec}, for example, take $\sim10$ seconds to compute
with a single CPU on modern computing hardware.  The quantum
trajectory Monte Carlo results in the same figure, meanwhile, take
$\sim10^4$ CPU hours to compute on similar hardware; the bulk of this
time is spent performing sparse matrix-vector multiplication, leaving
little room to further optimize serial runtime.  Parallelization can
reduce actual runtime of the Monte Carlo simulations to $\sim10$ hours
by running all trajectories at once, but at the cost of greatly
increasing computing resource requirements.  Though it may be possible
to further speed up quantum trajectory Monte Carlo simulations by
introducing new truncation schemes, any modifications
\begin{enumerate*}
\item should be made carefully to ensure that simulations still yield
  correct results, and
\item are unlikely to bridge the orders of magnitude in computing
  resource requirements.
\end{enumerate*}

\begin{figure}
  \centering
  \includegraphics{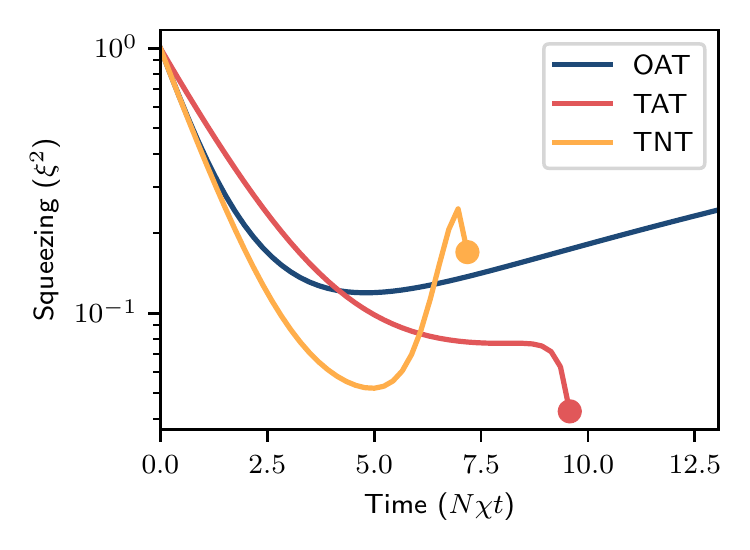}
  \caption{Spin squeezing of $N=10^4$ spins initially in $\ket\X$ with
    spontaneous decay, excitation, and dephasing of individual spins
    at rates $\gamma_-=\gamma_+=\gamma_\z=100\chi$.  Computed using
    the TST expansion in Eq.~\eqref{eq:TST} with $M=35$.  Solid
    circles mark the times at which the TST expansion gives an
    unphysical result with $\xi^2<0$.}
  \label{fig:decoherence_strong}
\end{figure}

The second advantage of the TST expansion is the capability to compute
spin correlators in strong-decoherence regimes of large systems that
are entirely inaccessible to other methods.  As an example, Figure
\ref{fig:decoherence_strong} shows squeezing of $N=10^4$ spins
initially in $\ket\X$, undergoing spontaneous decay, excitation, and
dephasing of individual spins at rates
$\gamma_-=\gamma_+=\gamma_\z=100\chi$.  The system size in these
simulations is too large for straightforward application of exact
methods for open quantum systems.  Quantum trajectory Monte Carlo
simulations, meanwhile, take a prohibitively long time to converge
with such strong decoherence due to the multiplicity of quantum
trajectories that require averaging.

The results in Figure \ref{fig:decoherence_strong} show that the TNT
model can generate more squeezing than the OAT or TAT models in the
presence of strong decoherence.  The better performance of TNT is in
part a consequence of the fact that TNT initially generates squeezing
at a faster rate than OAT or TAT, thereby allowing it to produce more
squeezing before the degrading effects of decoherence kick in.  We
corroborate this finding with quantum trajectory simulations of a
smaller system in Appendix \ref{sec:strong_dec}.  Strong-decoherence
computations of the sort used for Figure \ref{fig:decoherence_strong}
put lower bounds on theoretically achievable spin squeezing via TAT
with decoherence in Ref.~[\citenum{he2019engineering}], exemplifying a
concrete and practical application of the TST expansion and the
collective-spin structure constants calculated in this work.

%%%%%%%%%%%%%%%%%%%%%%%%%%%%%%%%%%%%%%%%%%%%%%%%%%%%%%%%%%%%%%%%%%%%%%
\section{Two-time correlation functions and out-of-time-ordered
  correlators}
\label{sec:multi_time}

As a final example of collective-spin physics that is numerically
accessible via the TST expansion of Heisenberg operators, we consider
the calculation of two-time correlation functions and
out-of-time-ordered correlators (OTOCs).  In particular, we consider
the effect of decoherence on short-time behavior of the two-time
connected correlator
\begin{align}
  C\p{t}
  \equiv \f1S\p{\bk{\hat S_+\p{t} \hat S_-\p{0}}
    - \bk{\hat S_+\p{t}}\bk{\hat S_-\p{0}}},
  \label{eq:two_time}
\end{align}
and the expectation value of a squared commutator,
\begin{align}
  D\p{t}
  \equiv \f1{S^2} \Bk{\sp{\hat S_+\p{t},\hat S_-\p{0}}_-^\dag
    \sp{\hat S_+\p{t},\hat S_-\p{0}}_-}_{\t{nn}},
  \label{eq:four_point}
\end{align}
in the context of the squeezing models in Section \ref{sec:squeezing}.
The subscript on $\bk{\cdot}_{\t{nn}}$ in Eq.~\eqref{eq:four_point}
stands for ``no noise'', and denotes a correlator computed without the
noise contributions $\hat\E_\O\p{t}$ to Heisenberg operators
$\hat\O\p{t}$.  While linear contributions from noise operators as
e.g.~in Eq.~\eqref{eq:two_time} always vanish under Markovian
decoherence (see Appendix \ref{sec:noise}), quadratic contributions
that would otherwise appear in Eq.~\eqref{eq:four_point} generally do
not\cite{blocher2019quantum}.  Determining the effect of these noise
terms generally requires making additional assumptions about the
environment, which would be a digression for the purposes of the
present work.  We therefore exclude these noise terms in
\eqref{eq:four_point} in order to keep our discussion simple and
general; see Ref.~[\citenum{blocher2019quantum}] for more detailed
discussions of noise terms and the quantum regression theorem
underlying the calculation of multi-time correlators.

In an equilibrium setting, correlation functions similar to that in
Eq.~\eqref{eq:two_time} contain information about the linear response
of Heisenberg operators to perturbations of a system; in a
non-equilibrium setting, they contribute to short-time linear response
(see Appendix \ref{sec:linear_response}).  Similar correlators have
made appearances as order parameters for diagnosing time-crystalline
phases of matter\cite{tucker2018shattered}.  Squared commutators such
as that in Eq.~\eqref{eq:four_point}, meanwhile, are commonly examined
for signatures of quantum chaos and information
scrambling\cite{maldacena2016bound, swingle2018unscrambling,
  garcia-mata2018chaos}.  In typical scenarios, such squared
commutators initially vanish by construction through a choice of
spatially separated operators.  Collective spin systems, however, have
no intrinsic notion of locality or spatial separation.  In our case,
therefore, with the choice of initial state
$\ket\X\propto\p{\ket\up+\ket\dn}^{\otimes N}$ we merely have
$D\p{0}\sim1/N$.

\begin{figure*}
  \centering
  \begin{minipage}{0.49\linewidth}
    \subfloat[Two-time correlator
    $C\p{t}\equiv\abs{C\p{t}}\exp\sp{i\phi\p{t}}$.
    \label{fig:two_time}]
    {\includegraphics{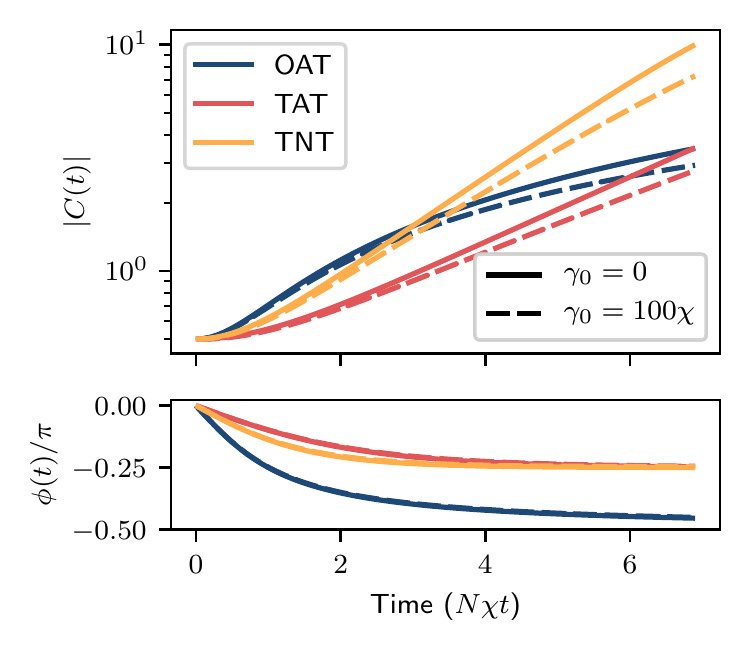}}
  \end{minipage}
  \begin{minipage}{0.49\linewidth}
    \subfloat[Squared commutator $D\p{t}$.
    \label{fig:four_point}]
    {\includegraphics{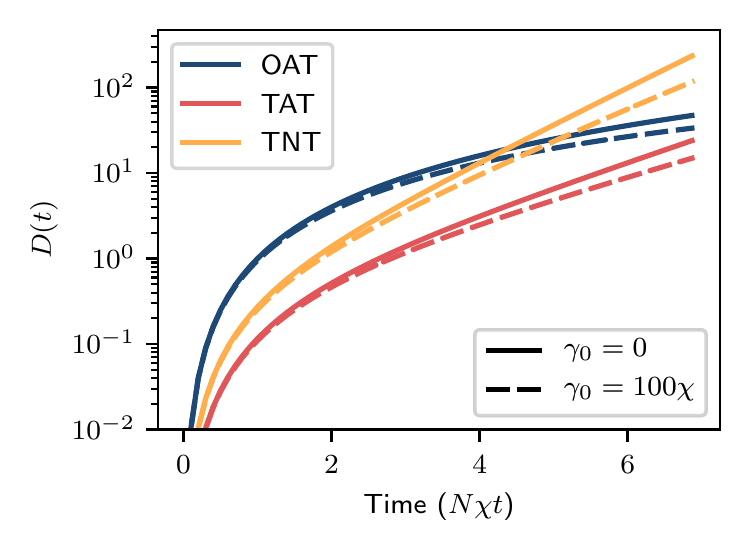}}
  \end{minipage}
  \caption{The two-time connected correlator $C\p{t}$ and squared
    commutator $D\p{t}$, respectively defined in
    Eqs.~\eqref{eq:two_time} and \eqref{eq:four_point}, for $N=10^4$
    spins initially in the polarized state
    $\ket\X\propto\p{\ket\up+\ket\dn}^{\otimes N}$ evolving under the
    squeezing Hamiltonians in Eqs.~\eqref{eq:OAT}--\eqref{eq:TNT}.
    Results are shown for both unitary dynamics (solid lines) and
    non-unitary dynamics with
    $\gamma_-=\gamma_+=\gamma_\z=\gamma_0=100\chi$ (dashed lines),
    computed using the TST expansion in Eq.~\eqref{eq:TST} with
    $M=20$.}
  \label{fig:multi_time}
\end{figure*}

Figure \ref{fig:multi_time} shows the behavior of $C\p{t}$ and
$D\p{t}$ for $N=10^4$ spins, initially in the state $\ket\X$, evolving
under the squeezing Hamiltonians in
Eqs.~\eqref{eq:OAT}--\eqref{eq:TNT} both with and without spontaneous
decay, excitation, and dephasing of individual spins at rates
$\gamma_-=\gamma_+=\gamma_\z=100\chi$.  In the case of unitary
evolution under OAT, we find that to an excellent approximation
$\abs{C\p{t}}$ takes the functional form
$f\p{t}=f\p{0}+aN\chi t+\p{bN\chi t}^2$ with $a\sim b\sim 1$, and with
a virtually perfect fit $D\p{t}=D\p{0}+\p{\sp{N+1}\chi t}^2$.  For
unitary evolution under TAT and TNT, we find that to an excellent
approximation both $\abs{C\p{t}}$ and $D\p{t}$ take the functional
form $f\p{t}=f\p{0}+a\sp{\exp\p{bN\chi t}-1}$ with $a\sim b\sim 1$.
As may be expected, the growth of $C\p{t}$ and $D\p{t}$ is generally
suppressed by decoherence.  Figure \ref{fig:multi_time} serves as an
example for the type of behavior that is accessible at short times
with the TST expansion.  These examples are straightforward to extend
to equilibrium settings and spin-boson systems.

%%%%%%%%%%%%%%%%%%%%%%%%%%%%%%%%%%%%%%%%%%%%%%%%%%%%%%%%%%%%%%%%%%%%%%
\section{Conclusions}

We have presented an efficient method for computing correlators at
short times in collective spin systems.  This method is based on
truncating a short-time expansion of Heisenberg operators, and can
access correlators on time scales that are relevant to metrological
applications such as spin squeezing.  In order to evaluate the
truncated short-time (TST) expansion of Heisenberg operators, we have
computed the structure constants of a collective spin operator
algebra, which we hope will empower future analytical and numerical
studies of collective spin systems.  Even though we considered only
non-equilibrium spin-squeezing processes in this work, our method can
be applied directly in an equilibrium setting, and is straightforward
to generalize to systems such as trapped ions and optical cavities
with collective spin-boson interactions.  In such contexts, our method
may be used to benchmark the short-time effects of decoherence, or
study the onset of quantum chaos and information scrambling.

\begin{acknowledgments}
  We acknowledge helpful discussions with Robert Lewis-Swan, Kris
  Tucker, and Colin Kennedy; as well as some technical contributions
  from Diego Barberena.  This work is supported by the Air Force
  Office of Scientific Research (AFOSR) grant FA9550-18-1-0319; the
  Defense Advanced Research Projects Agency (DARPA) and Army Research
  Office (ARO) grant W911NF-16-1-0576 and W911NF-19-1-0210; the
  National Science Foundation (NSF) grant PHY-1820885; JILA-NSF grant
  PFC-1734006; and the National Institute of Standards and Technology
  (NIST).
\end{acknowledgments}

\onecolumngrid
\appendix

%%%%%%%%%%%%%%%%%%%%%%%%%%%%%%%%%%%%%%%%%%%%%%%%%%%%%%%%%%%%%%%%%%%%%%
\section{Basic spin operator identities}
\label{sec:identities}

The appendices in this work make ubiquitous use of various spin
operator identities; we collect and derive some basic identities here
for reference.  Note that despite the working definition of collective
spin operators from $S_\a=\sum_js_\a^{(j)}$, the identities we will
derive involving only collective spin operators apply just as well to
large-spin operators that cannot be expressed as the sum of individual
spin-1/2 operators.  The elementary commutation relations between spin
operators are, with $\bmu\equiv-\mu\in\set{+1,-1}$ for brevity,
\begin{align}
  \sp{s_\z^{(j)},s_\mu^{(k)}}_-
  &= \delta_{jk} \mu s_\mu^{(j)},
  &
  \sp{S_\z,s_\mu^{(j)}}_-
  &= \sp{s_\z^{(j)},S_\mu}_- = \mu s_\mu^{(j)},
  &
  \sp{S_\z,S_\mu}_-
  &= \mu S_\mu,
  \label{eq:comm_z_base} \\
  \sp{s_\mu^{(j)},s_\bmu^{(k)}}_-
  &= \delta_{jk} 2 \mu s_\z^{(j)},
  &
  \sp{S_\mu,s_\bmu^{(j)}}_-
  &= \sp{s_\mu^{(j)},S_\bmu}_- = 2 \mu s_\z^{(j)},
  &
  \sp{S_\mu,S_\bmu}_-
  &= 2 \mu S_\z.
  \label{eq:comm_mu_base}
\end{align}
These relations can be used to inductively compute identities
involving powers of collective spin operators.  By pushing through one
spin operator at a time, we can find
\begin{align}
  \p{\mu S_\z}^m s_\mu^{(j)}
  = \p{\mu S_\z}^{m-1} s_\mu^{(j)} \p{1 + \mu S_\z}
  = \p{\mu S_\z}^{m-2} s_\mu^{(j)} \p{1 + \mu S_\z}^2
  = \cdots
  = s_\mu^{(j)} \p{1 + \mu S_\z}^m,
  \label{eq:push_z_mu_Ss}
\end{align}
and
\begin{align}
  \mu s_\z^{(j)} S_\mu^m
  = S_\mu \mu s_\z^{(j)} S_\mu^{m-1} + s_\mu^{(j)} S_\mu^{m-1}
  = \cdots
  = S_\mu^m \mu s_\z^{(j)} + ms_\mu^{(j)} S_\mu^{m-1},
  \label{eq:push_z_mu_sS}
\end{align}
where we will generally find it nicer to express results in terms of
$\mu s_\z^{(j)}$ and $\mu S_\z$ rather than $s_\z^{(j)}$ and $S_\z$.
Summing over the single-spin index $j$ in both of the cases above
gives us the purely collective-spin versions of these identities:
\begin{align}
  \p{\mu S_\z}^m S_\mu = S_\mu \p{1 + \mu S_\z}^m,
  &&
  \mu S_\z S_\mu^m = S_\mu^m \p{m + \mu S_\z},
  \label{eq:push_z_mu_single}
\end{align}
where we can repeat the process of pushing through individual $S_\z$
operators $\ell$ times to get
\begin{align}
  \p{\mu S_\z}^\ell S_\mu^m
  = \p{\mu S_\z}^{\ell-1} S_\mu^m \p{m + \mu S_\z}
  = \p{\mu S_\z}^{\ell-2} S_\mu^m \p{m + \mu S_\z}^2
  = \cdots
  = S_\mu^m \p{m + \mu S_\z}^\ell.
  \label{eq:push_z_mu}
\end{align}
Multiplying \eqref{eq:push_z_mu} through by $\p{\mu\nu}^\ell$ (for
$\nu\in\set{+1,-1}$) and taking its Hermitian conjugate, we can say
that more generally
\begin{align}
  \p{\nu S_\z}^\ell S_\mu^m
  = S_\mu^m \p{\mu\nu m+\nu S_\z}^\ell,
  &&
  S_\mu^m \p{\nu S_\z}^\ell
  = \p{-\mu\nu m+\nu S_\z}^\ell S_\mu^m.
\end{align}
Finding commutation relations between powers of transverse spin
operators, i.e.~$S_\mu$ and $S_\bmu$, turns out to be considerably
more difficult than the cases we have worked out thus far.  We
therefore save this work for Appendix \ref{sec:comm_transverse}.

%%%%%%%%%%%%%%%%%%%%%%%%%%%%%%%%%%%%%%%%%%%%%%%%%%%%%%%%%%%%%%%%%%%%%%
\section{Commutation relations between powers of transverse spin
  operators}
\label{sec:comm_transverse}

To find commutation relations between powers of transverse collective
spin operators, we first compute
\begin{align}
  S_\mu^m s_\bmu^{(j)}
  &= S_\mu^{m-1}s_\bmu^{(j)} S_\mu
  + S_\mu^{m-1} 2\mu s_\z^{(j)} \\
  &= S_\mu^{m-2} s_\bmu^{(j)} S_\mu^2
  + S_\mu^{m-2} 2\mu s_\z^{(j)} S_\mu
  + S_\mu^{m-1} 2\mu s_\z^{(j)} \\
  &= s_\bmu^{(j)} S_\mu^m
  + \sum_{k=0}^{m-1} S_\mu^k 2\mu s_\z^{(j)} S_\mu^{m-k-1}
  \label{eq:push_mu_Ss_start}.
\end{align}
While \eqref{eq:push_mu_Ss_start} gives us the commutator
$\sp{S_\mu^m,s_\bmu^{(j)}}_-$, we would like to enforce an ordering on
products of spin operators, which will ensure that we only keep track
of operators that are linearly independent.  We choose (for now) to
impose an ordering with all $s_\bmu^{(j)}$ operators on the left, and
all $s_\z^{(j)}$ operators on the right.  Such an ordering will prove
convenient for the calculations in this section\footnote{In
  retrospect, it may have been nicer to push all $s_\mu^{(j)}$
  operators to the right throughout these calculations, due to the
  enhanced symmetry that expressions would have with respect to
  Hermitian conjugation.  In any case, we provide the final result of
  this section in both ordering conventions, and therefore feel no
  need to reproduce these calculations with a different ordering of
  spin operators.}.  This choice of ordering compels us to expand
\begin{align}
  \sum_{k=0}^{m-1} S_\mu^k 2\mu s_\z^{(j)} S_\mu^{m-k-1}
  &= \sum_{k=0}^{m-1} S_\mu^k
  \sp{2\p{m-k-1} s_\mu^{(j)} S_\mu^{m-k-2}
    + S_\mu^{m-k-1} 2\mu s_\z^{(j)}} \\
  &= m \p{m-1} s_\mu^{(j)} S_\mu^{m-2}
  + m S_\mu^{m-1} 2\mu s_\z^{(j)},
\end{align}
which implies
\begin{align}
  S_\mu^m s_\bmu^{(j)}
  = s_\bmu^{(j)} S_\mu^m + m \p{m-1} s_\mu^{(j)} S_\mu^{m-2}
  + m S_\mu^{m-1} 2\mu s_\z^{(j)},
  \label{eq:push_mu_Ss}
\end{align}
and in turn
\begin{align}
  S_\mu^m S_\bmu = S_\bmu S_\mu^m
  + m S_\mu^{m-1} \p{m - 1 + 2\mu S_\z}.
  \label{eq:push_mu_single}
\end{align}
As the next logical step, we take on the task of computing
\begin{align}
  S_\mu^m S_\bmu^n
  = S_\mu^{m-1} S_\bmu^n S_\mu
  + n \sp{S_\mu^{m-1} S_\bmu^{n-1} \p{1 - n + 2\mu S_\z}}
  = S_\bmu^n S_\mu^m
  + n \sum_{k=0}^{m-1} S_\mu^{m-k-1} S_\bmu^{n-1}
  \p{1 - n + 2\mu S_\z} S_\mu^k,
\end{align}
which implies
\begin{align}
  \sp{S_\mu^m, S_\bmu^n}_-
  = C_{mn;\mu}
  \equiv n \sum_{k=0}^{m-1} S_\mu^{m-k-1} S_\bmu^{n-1}
  \p{1 - n + 2\mu S_\z} S_\mu^k.
\end{align}
We now need to rearrange the operators in $C_{mn;\mu}$ into a standard
order, which means pushing all $S_\z$ operators to the right and, for
the purposes of this calculation, all $S_\bmu$ operators to the left.
We begin by pushing $S_\mu^k$ to the left of $S_\z$, which takes
$2\mu S_\z\to 2\mu S_\z+2k$, and then push $S_\mu^{m-k-1}$ to the
right of $S_\bmu^{n-1}$, giving us
\begin{align}
  C_{mn;\mu}
  &= n \sum_{k=0}^{m-1}
  \p{S_\bmu^{n-1} S_\mu^{m-k-1} + C_{m-k-1,n-1;\mu}} S_\mu^k
  \p{2k + 1 - n + 2\mu S_\z} \\
  &= D_{mn;\mu}
  + n \sum_{k=0}^{m-2} C_{m-k-1,n-1;\mu}
  S_\mu^k \p{2k + 1 - n + 2\mu S_\z},
  \label{eq:C_mn}
\end{align}
where we have dropped the last ($k=m-1$) term in the remaining sum
because $C_{m-k-1,n-1;\mu}=0$ if $k=m-1$, and
\begin{align}
  D_{mn;\mu}
  \equiv mn S_\bmu^{n-1} S_\mu^{m-1} \p{m - n + 2\mu S_\z}.
  \label{eq:D_mn}
\end{align}
To our despair, we have arrived in \eqref{eq:C_mn} at a {\it
  recursive} formula for $C_{mn;\mu}$.  Furthermore, we have not even
managed to order all spin operators, as $C_{m-k-1,n-1;\mu}$ contains
$S_\z$ operators that are to the left of $S_\mu^k$.  To sort all spin
operators once and for all, we define
\begin{align}
  C_{mn;\mu}^{(k)} \equiv C_{m-k,n;\mu} S_\mu^k,
  &&
  D_{mn;\mu}^{(k)} \equiv D_{m-k,n;\mu} S_\mu^k,
\end{align}
which we can expand as
\begin{align}
  D_{mn;\mu}^{(k)}
  &= \p{m-k}n S_\bmu^{n-1} S_\mu^{m-k-1}
  \p{m-k-n+2\mu S_\z} S_\mu^k \\
  &= \p{m-k}n S_\bmu^{n-1} S_\mu^{m-1} \p{k+m-n+2\mu S_\z},
  \label{eq:D_mn_k}
\end{align}
and
\begin{align}
  C_{mn;\mu}^{(k)}
  &= D_{m-k,n;\mu} S_\mu^k + n \sum_{j=0}^{m-k-2}
  C_{m-k-j-1,n-1;\mu} S_\mu^j \p{2j+1-n+2\mu S_\z} S_\mu^k \\
  &= D_{mn;\mu}^{(k)} + n \sum_{j=0}^{m-k-2}
  C_{m-k-j-1,n-1;\mu} S_\mu^{j+k} \p{2j+2k+1-n+2\mu S_\z} \\
  &= D_{mn;\mu}^{(k)} + n \sum_{j=0}^{m-k-2}
  C_{m-1,n-1;\mu}^{(k+j)} \p{2\sp{j+k}+1-n+2\mu S_\z} \\
  &= D_{mn;\mu}^{(k)} + n \sum_{j=k}^{m-2}
  C_{m-1,n-1;\mu}^{(j)} \p{2j+1-n+2\mu S_\z}.
  \label{eq:C_mn_k}
\end{align}
While the resulting expression in \eqref{eq:C_mn_k} strongly resembles
that in \eqref{eq:C_mn}, there is one crucial difference: all spin
operators in \eqref{eq:C_mn_k} have been sorted into a standard order.
We can now repeatedly substitute $C_{mn;\mu}^{(k)}$ into itself, each
time decreasing $m$ and $n$ by 1, until one of $m$ or $n$ reaches
zero.  Such repeated substitution yields the expansion
\begin{align}
  C_{mn;\mu}
  = C_{mn;\mu}^{(0)}
  = D_{mn;\mu}
  + \sum_{p=1}^{\min\set{m,n}-1} E_{mn;\mu}^{(p)},
  \label{eq:C_mn_E}
\end{align}
where the first two terms in the sum over $p$ are
\begin{align}
  E_{mn;\mu}^{(1)}
  &= n \sum_{k=0}^{m-2} D_{m-1,n-1;\mu}^{(k)} \p{2k+1-n+2\mu S_\z}, \\
  E_{mn;\mu}^{(2)}
  &= n \sum_{k_1=0}^{m-2} \p{n-1} \sum_{k_2=k_1}^{m-3}
  D_{m-2,n-2;\mu}^{(k_2)} \p{2k_2+2-n+2\mu S_\z} \p{2k_1+1-n+2\mu S_\z},
\end{align}
and more generally for $p>1$,
\begin{align}
  E_{mn;\mu}^{(p)}
  = \f{n!}{\p{n-p}!}
  \sum_{k_1=0}^{m-2} \sum_{k_2=k_1}^{m-3} \cdots\sum_{k_p=k_{p-1}}^{m-p-1}
  D_{m-p,n-p;\mu}^{(k_p)} \prod_{j=1}^p \p{2k_j+j-n+2\mu S_\z}.
  \label{eq:E_mn_p}
\end{align}
In principle, the expressions in \eqref{eq:D_mn}, \eqref{eq:D_mn_k},
\eqref{eq:C_mn_E}, and \eqref{eq:E_mn_p} suffice to evaluate the
commutator $\sp{S_\mu^m,S_\bmu^n}_- = C_{mn;\mu}$, but this result is
-- put lightly -- quite a mess: the expression for $E_{mn;\mu}^{(p)}$
in \eqref{eq:E_mn_p} involves a sum over $p$ mutually dependent
intermediate variables, each term of which additionally contains a
product of $p$ factors.  We therefore devote the rest of this section
to simplifying our result for the commutator
$\sp{S_\mu^m,S_\bmu^n}_-$.

Observing that in \eqref{eq:E_mn_p} we always have
$0\le k_1\le k_2\le\cdots\le k_p\le m-p-1$, we can rearrange the order
of the sums and relabel $k_p\to\ell$ to get
\begin{align}
  E_{mn;\mu}^{(p)}
  = \f{n!}{\p{n-p}!}
  \sum_{\ell=0}^{m-p-1} D_{m-p,n-p;\mu}^{(\ell)} \p{2\ell+F_{np;\mu}}
  \sum_{\p{\v k,p-1,\ell}} \prod_{j=1}^{p-1} \p{2k_{p-j}-j+F_{np;\mu}},
  \label{eq:E_mn_p_sum}
\end{align}
where for shorthand we define
\begin{align}
  F_{np;\mu} \equiv p - n + 2\mu S_\z,
  &&
  \sum_{\p{\v k,q,\ell}} X \equiv
  \sum_{k_1=0}^\ell \sum_{k_2=k_1}^\ell
  \cdots \sum_{k_q=k_{q-1}}^\ell X.
\end{align}
We now further define
\begin{align}
  f_{np\ell;\mu}\p{k,q} \equiv \p{\ell-k+q} \p{\ell+k-q+F_{np;\mu}},
\end{align}
and evaluate sums successively over $k_{p-1},k_{p-2},\cdots,k_1$,
finding
\begin{align}
  \sum_{\p{\v k,p-1,\ell}} \prod_{j=1}^{p-1} \p{2k_{p-j}-j+F_{np;\mu}}
  &= \sum_{\p{\v k,p-2,\ell}}
  \prod_{j=2}^{p-1} \p{2k_{p-j}-j+F_{np;\mu}}
  f_{np\ell;\mu}\p{k_{p-2},1} \\
  &= \f1{\p{r-1}!} \sum_{\p{\v k,p-r,\ell}}
  \prod_{j=r}^{p-1} \p{2k_{p-j}-j+F_{np;\mu}}
  \prod_{q=1}^{r-1} f_{np\ell;\mu}\p{k_{p-r},q} \\
  &= \f1{\p{p-1}!} \prod_{q=1}^{p-1} f_{np\ell;\mu}\p{0,q} \\
  &= { \ell + p - 1 \choose p - 1 }
  \prod_{q=1}^{p-1} \p{\ell-q+F_{np;\mu}}.
\end{align}
Substitution of this result together with $D_{m-p,n-p;\mu}^{(\ell)}$
using \eqref{eq:D_mn_k} into \eqref{eq:E_mn_p_sum} then gives us
\begin{align}
  E_{mn;\mu}^{(p)}
  = \f{n!}{\p{n-p-1}!} S_\bmu^{n-p-1} S_\mu^{m-p-1} G_{mnp;\mu}
\end{align}
with
\begin{align}
  G_{mnp;\mu}
  &\equiv \sum_{\ell=0}^{m-p-1} { \ell + p - 1 \choose p - 1 }
  \p{m-p-\ell} \p{\ell+m-p+F_{np;\mu}}
  \p{2\ell + F_{np;\mu}}
  \prod_{q=1}^{p-1} \p{\ell-q+F_{np;\mu}} \\
  &= { m \choose p + 1 } \prod_{q=0}^p \p{m-p-q+F_{np;\mu}}.
\end{align}
We can further simplify
\begin{align}
  \prod_{q=0}^p \p{m-p-q+F_{np;\mu}}
  = \prod_{q=0}^p \p{m-n-q+2\mu S_\z}
  = \sum_{q=0}^{p+1} \p{-1}^{p+1-q}
  { p+1 \brack q } \p{m-n+2\mu S_\z}^q,
\end{align}
where ${ p \brack q }$ is an unsigned Stirling number of the first
kind, and finally
\begin{align}
  \sum_{q=0}^p \p{-1}^{p-q} { p \brack q } \p{m-n+2\mu S_\z}^q
  &= \sum_{q=0}^p \p{-1}^{p-q} { p \brack q } \sum_{\ell=0}^q
  { q \choose \ell } \p{m-n}^{q-\ell} \p{2\mu S_\z}^\ell \\
  &= \sum_{\ell=0}^p 2^\ell \sum_{q=\ell}^p \p{-1}^{p-q}
  { p \brack q } { q \choose \ell } \p{m-n}^{q-\ell} \p{\mu S_\z}^\ell.
\end{align}
Putting everything together, we finally have
\begin{align}
  E_{mn;\mu}^{(p-1)}
  = p! { m \choose p } { n \choose p }
  S_\bmu^{n-p} S_\mu^{m-p}
  \sum_{\ell=0}^p \epsilon_{mn}^{p\ell} \p{\mu S_\z}^\ell,
\end{align}
with
\begin{align}
  \epsilon_{mn}^{p\ell}
  \equiv 2^\ell \sum_{q=\ell}^p \p{-1}^{p-q}
  { p \brack q } { q \choose \ell } \p{m-n}^{q-\ell},
\end{align}
where in this final form $E_{mn;\mu}^{(0)} = D_{mn;\mu}$, which
together with the expansion for $C_{mn;\mu}$ in \eqref{eq:C_mn_E}
implies that
\begin{align}
  \sp{S_\mu^m, S_\bmu^n}_-
  = \sum_{p=1}^{\min\set{m,n}}
  p! { m \choose p } { n \choose p } S_\bmu^{n-p} S_\mu^{m-p}
  \sum_{\ell=0}^p \epsilon_{mn}^{p\ell} \p{\mu S_\z}^\ell,
  \label{eq:comm_mu}
\end{align}
and
\begin{align}
  S_\mu^m S_\bmu^n
  = \sum_{p=0}^{\min\set{m,n}}
  p! { m \choose p } { n \choose p } S_\bmu^{n-p} S_\mu^{m-p}
  \sum_{\ell=0}^p \epsilon_{mn}^{p\ell} \p{\mu S_\z}^\ell.
  \label{eq:push_mu_bmu}
\end{align}
If we wish to order products of collective spin operators with $S_\z$
in between $S_\bmu$ and $S_\mu$, then
\begin{align}
  S_\mu^m S_\bmu^n
  = \sum_{p=0}^{\min\set{m,n}} p! { m \choose p } { n \choose p }
  S_\bmu^{n-p} Z_{mn;\bmu}^{(p)} S_\mu^{m-p},
\end{align}
where
\begin{align}
  Z_{mn;\bmu}^{(p)}
  \equiv \sum_{\ell=0}^p \epsilon_{mn}^{p\ell}
  \p{-\sp{m-p} + \mu S_\z}^\ell
  = \sum_{q=0}^p \zeta_{mn}^{pq} \p{\bmu S_\z}^q,
  \label{eq:Z_mnp}
\end{align}
with
\begin{align}
  \zeta_{mn}^{pq}
  \equiv \sum_{\ell=q}^p \epsilon_{mn}^{p\ell}
  { \ell \choose q } \p{-1}^\ell \p{m-p}^{\ell-q}
  = \p{-1}^p 2^q \sum_{s=q}^p
  { p \brack s } { s \choose q } \p{m+n-2p}^{s-q}.
  \label{eq:zeta_mnpq}
\end{align}
Here ${ p \brack s }$ is an unsigned Stirling number of the first
kind.

%%%%%%%%%%%%%%%%%%%%%%%%%%%%%%%%%%%%%%%%%%%%%%%%%%%%%%%%%%%%%%%%%%%%%%
\section{Product of arbitrary ordered collective spin operators}
\label{sec:general_product}

The most general product of collective spin operators that we need to
compute is
\begin{align}
  \S^{pqr}_{\ell mn;\mu}
  = S_\mu^p \p{\mu S_\z}^q S_\bmu^r
  S_\mu^\ell \p{\mu S_\z}^m S_\bmu^n
  = \sum_{k=0}^{\min\set{r,\ell}} k! { r \choose k } { \ell \choose k }
  S_\mu^{p+\ell-k} \tilde Z_{qr\ell m;\mu}^{(k)} S_\bmu^{r+n-k},
  \label{eq:general_product}
\end{align}
where
\begin{align}
  \tilde Z_{qr\ell m;\mu}^{(k)}
  &\equiv \p{\ell-k+\mu S_\z}^q
  Z_{r\ell;\mu}^{(k)} \p{r-k+\mu S_\z}^m \\
  &= \sum_{a=0}^k \zeta_{r\ell}^{ka}
  \sum_{b=0}^q \p{\ell-k}^{q-b} { q \choose b }
  \sum_{c=0}^m \p{r-k}^{m-c} { m \choose c }
  \p{\mu S_\z}^{a+b+c},
\end{align}
is defined in terms of $Z_{r\ell;\mu}^{(k)}$ and $\zeta_{r\ell}^{ka}$
as respectively given in \eqref{eq:Z_mnp} and \eqref{eq:zeta_mnpq}.
The (anti-)commutator of two ordered products of collective spin
operators is then simply
\begin{align}
  \sp{S_\mu^p \p{\mu S_\z}^q S_\bmu^r,
    S_\mu^\ell \p{\mu S_\z}^m S_\bmu^n}_\pm
  = \S^{pqr}_{\ell mn;\mu} \pm \S^{\ell mn}_{pqr;\mu}.
\end{align}

%%%%%%%%%%%%%%%%%%%%%%%%%%%%%%%%%%%%%%%%%%%%%%%%%%%%%%%%%%%%%%%%%%%%%%
\section{Sandwich identities for single-spin decoherence calculations}
\label{sec:sandwich_single}

In this section we derive several identities that will be necessary
for computing the effects of single-spin decoherence on ordered
products of collective spin operators, i.e.~on operators of the form
$S_\mu^\ell \p{\mu S_\z}^m S_\bmu^n$.  These identities all involve
sandwiching a collective spin operator between operators that act on
individual spins only, and summing over all individual spin indices.
Our general strategy will be to use commutation relations to push
single-spin operators together, and then evaluate the sum to arrive at
an expression involving only collective spin operators.

We first compute sums of single-spin operators sandwiching
$\p{\mu S_\z}^m$, when necessary making use of the identity in
\eqref{eq:push_z_mu_Ss}.  The unique cases up to Hermitian conjugation
are, for $S\equiv N/2$ and $\mu,\nu\in\set{+1,-1}$,
\begin{align}
  \sum_j s_\z^{(j)} \p{\mu S_\z}^m s_\z^{(j)}
  &= \sum_j s_\z^{(j)} s_\z^{(j)} \p{\mu S_\z}^m
  = \f14 \sum_j \1_j \p{\mu S_\z}^m
  = \f12 S \p{\mu S_\z}^m, \\
  \sum_j s_\z^{(j)} \p{\mu S_\z}^m s_\nu^{(j)}
  &= \p{\mu S_\z}^m \sum_j s_\z^{(j)} s_\nu^{(j)}
  = \f12 \p{\mu S_\z}^m \nu S_\nu
  = \f12 \nu S_\nu \p{\mu\nu+\mu S_\z}^m, \\
  \sum_j s_\nu^{(j)} \p{\mu S_\z}^m s_\nu^{(j)}
  &= \sum_j s_\nu^{(j)} s_\nu^{(j)} \p{\mu\nu+\mu S_\z}^m
  = 0, \\
  \sum_j s_\bnu^{(j)} \p{\mu S_\z}^m s_\nu^{(j)}
  &= \sum_j s_\bnu^{(j)} s_\nu^{(j)} \p{\mu\nu+\mu S_\z}^m
  = \p{S-\nu S_\z} \p{\mu\nu+\mu S_\z}^m.
\end{align}
We are now equipped to derive similar identities for more general
collective spin operators.  Making heavy use of identities
\eqref{eq:push_z_mu_sS} and \eqref{eq:push_mu_Ss} to push single-spin
operators through transverse collective-spin operators, we again work
through all combinations that are unique up to Hermitian conjugation,
finding
\begin{align}
  \sum_j s_\z^{(j)} S_\mu^\ell \p{\mu S_\z}^m S_\bmu^n s_\z^{(j)}
  &= \f12 \p{S-\ell-n} S_\mu^\ell \p{\mu S_\z}^m S_\bmu^n
  + \ell n S_\mu^{\ell-1} \p{S+\mu S_\z}
  \p{-1+\mu S_\z}^m S_\bmu^{n-1},
  \label{eq:san_z_z} \allowdisplaybreaks \\
  \sum_j s_\z^{(j)} S_\mu^\ell \p{\mu S_\z}^m S_\bmu^n s_\mu^{(j)}
  &= \f12 \mu S_\mu^{\ell+1} \p{1+\mu S_\z}^m S_\bmu^n
  - \mu n \p{S-\ell-\f12\sp{n-1}} S_\mu^\ell
  \p{\mu S_\z}^m S_\bmu^{n-1} \notag \\
  &\qquad - \mu\ell n\p{n-1} S_\mu^{\ell-1}
  \p{S+\mu S_\z} \p{-1+\mu S_\z}^m S_\bmu^{n-2},
  \label{eq:san_z_mu} \allowdisplaybreaks \\
  \sum_j s_\z^{(j)} S_\mu^\ell \p{\mu S_\z}^m S_\bmu^n s_\bmu^{(j)}
  &= -\f12 \mu S_\mu^\ell \p{\mu S_\z}^m S_\bmu^{n+1}
  + \mu \ell S_\mu^{\ell-1} \p{S+\mu S_\z} \p{-1+\mu S_\z}^m S_\bmu^n,
  \label{eq:san_z_bmu} \allowdisplaybreaks \\
  \sum_j s_\mu^{(j)} S_\mu^\ell \p{\mu S_\z}^m S_\bmu^n s_\mu^{(j)}
  &= n S_\mu^{\ell+1} \p{\mu S_\z}^m S_\bmu^{n-1}
  - n\p{n-1} S_\mu^\ell \p{S+\mu S_\z} \p{-1+\mu S_\z}^m S_\bmu^{n-2},
  \label{eq:san_mu_mu} \allowdisplaybreaks \\
  \sum_j s_\mu^{(j)} S_\mu^\ell \p{\mu S_\z}^m S_\bmu^n s_\bmu^{(j)}
  &= S_\mu^\ell \p{S+\mu S_\z}\p{-1+\mu S_\z}^m S_\bmu^n,
  \label{eq:san_mu_bmu} \allowdisplaybreaks \\
  \sum_j s_\bmu^{(j)} S_\mu^\ell \p{\mu S_\z}^m S_\bmu^n s_\mu^{(j)}
  &= S_\mu^\ell \p{S - \ell - n - \mu S_\z}
  \p{1+\mu S_\z}^m S_\bmu^n
  + \ell n \p{2S - \ell - n + 2}
  S_\mu^{\ell-1} \p{\mu S_\z}^m S_\bmu^{n-1} \notag \\
  &\qquad + \ell n \p{\ell-1} \p{n-1} S_\mu^{\ell-2} \p{S+\mu S_\z}
  \p{-1+\mu S_\z}^m S_\bmu^{n-2}.
  \label{eq:san_bmu_mu}
\end{align}

%%%%%%%%%%%%%%%%%%%%%%%%%%%%%%%%%%%%%%%%%%%%%%%%%%%%%%%%%%%%%%%%%%%%%%
\section{Uncorrelated, permutationally-symmetric single-spin
  decoherence}
\label{sec:decoherence_single}

In this section we work out the effects of permutationally-symmetric
decoherence of individual spins on collective spin operators of the
form $S_\mu^\ell \p{\mu S_\z}^m S_\bmu^n$.  For compactness, we define
\begin{align}
  \D\p{g} \O
  \equiv \D\p{\set{g^{(j)}:j=1,2,\cdots,N}} \O
  = \sum_j\p{{g^{(j)}}^\dag \O g^{(j)}
    - \f12\sp{{g^{(j)}}^\dag g^{(j)}, \O}_+},
\end{align}
where $g$ is an operator that acts on a single spin, $g^{(j)}$ is an
operator that acts with $g$ on spin $j$ and trivially on all other
spins, and $N$ is the total number of spins.

%%%%%%%%%%%%%%%%%%%%%%%%%%%%%%%%%%%%%%%%%%%%%%%%%%%%%%%%%%%%%%%%%%%%%%
\subsection{Decay-type decoherence}
\label{sec:decay_single}

The effect of decoherence via uncorrelated decay ($\mu=-1$) or
excitation ($\mu=1$) of individual spins is described by
\begin{align}
  \D\p{s_\mu} \O
  = \sum_j\p{s_\bmu^{(j)} \O s_\mu^{(j)}
    - \f12\sp{s_\bmu^{(j)} s_\mu^{(j)},\O}_+}
  = \sum_j s_\bmu^{(j)} \O s_\mu^{(j)}
  - S \O + \f{\mu}{2} \sp{S_\z, \O}_+.
\end{align}
In order to determine the effect of this decoherence on general
collective spin operators, we expand the anti-commutator
\begin{align}
  \sp{S_\z, S_\mu^\ell \p{\mu S_\z}^m S_\bmu^n}_+
  = S_\z S_\mu^\ell \p{\mu S_\z}^m S_\bmu^n
  + S_\mu^\ell \p{\mu S_\z}^m S_\bmu^n S_\z
  = \mu S_\mu^\ell\p{\ell+n+2\mu S_\z} \p{\mu S_\z}^m S_\bmu^n,
\end{align}
which implies, using \eqref{eq:san_mu_bmu},
\begin{align}
  \D\p{s_\bmu} \p{S_\mu^\ell \p{\mu S_\z}^m S_\bmu^n}
  = S_\mu^\ell \p{S+\mu S_\z}\p{-1+\mu S_\z}^m S_\bmu^n
  - S_\mu^\ell\sp{S + \f12\p{\ell+n} + \mu S_\z}
  \p{\mu S_\z}^m S_\bmu^n,
  \label{eq:decay_diff}
\end{align}
and, using \eqref{eq:san_bmu_mu},
\begin{align}
  \D\p{s_\mu} \p{S_\mu^\ell \p{\mu S_\z}^m S_\bmu^n}
  &= S_\mu^\ell \p{S - \ell - n - \mu S_\z} \p{1+\mu S_\z}^m S_\bmu^n
  - S_\mu^\ell\sp{S - \f12\p{\ell+n} - \mu S_\z}
  \p{\mu S_\z}^m S_\bmu^n \notag \\
  &\qquad + \ell n \p{2S - \ell - n + 2}
  S_\mu^{\ell-1} \p{\mu S_\z}^m S_\bmu^{n-1} \notag \\
  &\qquad + \ell n \p{\ell-1} \p{n-1} S_\mu^{\ell-2} \p{S + \mu S_\z}
  \p{-1+\mu S_\z}^m S_\bmu^{n-2}.
  \label{eq:decay_same}
\end{align}
Decoherence via jump operators $s_\bmu^{(j)}$ only couples operators
$S_\mu^\ell \p{\mu S_\z}^m S_\bmu^n$ to operators
$S_\mu^\ell \p{\mu S_\z}^{m'} S_\bmu^n$ with $m'\le m$.  Decoherence
via jump operators $s_\mu^{(j)}$, meanwhile, makes operators
$S_\mu^\ell \p{\mu S_\z}^m S_\bmu^n$ ``grow'' in $m$ through the last
term in \eqref{eq:decay_same}, although the sum $\ell+m+n$ does not
grow.

%%%%%%%%%%%%%%%%%%%%%%%%%%%%%%%%%%%%%%%%%%%%%%%%%%%%%%%%%%%%%%%%%%%%%%
\subsection{Dephasing}
\label{sec:dephasing_single}

The effect of decoherence via single-spin dephasing is described by
\begin{align}
  \D\p{s_\z} \O
  = \sum_j\p{s_\z^{(j)} \O s_\z^{(j)}
    - \f12\sp{s_\z^{(j)} s_\z^{(j)},\O}_+}
  = \sum_j s_\z^{(j)} \O s_\z^{(j)} - \f12 S \O.
\end{align}
From \eqref{eq:san_z_z}, we then have
\begin{align}
  \D\p{s_\z} \p{S_\mu^\ell \p{\mu S_\z}^m S_\bmu^n}
  = -\f12\p{\ell+n} S_\mu^\ell \p{\mu S_\z}^m S_\bmu^n
  + \ell n S_\mu^{\ell-1} \p{S + \mu S_\z}
  \p{-1 + \mu S_\z}^m S_\bmu^{n-1}.
\end{align}
Decoherence via single-spin dephasing makes operators
$S_\mu^\ell \p{\mu S_\z}^m S_\bmu^n$ ``grow'' in $m$, although the sum
$\ell+m+n$ does not grow.

%%%%%%%%%%%%%%%%%%%%%%%%%%%%%%%%%%%%%%%%%%%%%%%%%%%%%%%%%%%%%%%%%%%%%%
\subsection{The general case}
\label{sec:general_single}

The most general type of single-spin decoherence is described by
\begin{align}
  \D\p{g} \O
  = \sum_j\p{{g^{(j)}}^\dag \O g^{(j)}
    - \f12\sp{{g^{(j)}}^\dag g^{(j)}, \O}_+},
  &&
  g \equiv g_\z s_\z + g_+ s_+ + g_- s_-.
  \label{eq:D_general_single}
\end{align}
To simplify \eqref{eq:D_general_single}, we expand
\begin{align}
  g^\dag \O g
  = \abs{g_\z}^2 s_\z \O s_\z
  + \sum_\mu \p{\abs{g_\mu}^2 s_\bmu \O s_\mu
    + g_\bmu^* g_\mu s_\mu \O s_\mu
    + g_\z^* g_\mu s_\z \O s_\mu
    + g_\bmu^* g_\z s_\mu \O s_\z},
\end{align}
and
\begin{align}
  g^\dag g
  = \f14 \abs{g_\z}^2
  + \f12 \sum_\mu \sp{\abs{g_\mu}^2 \p{1-2\mu s_\z}
    + \mu \p{g_\z^*g_\mu - g_\bmu^*g_\z} s_\mu},
\end{align}
which implies
\begin{align}
  \D\p{g} \O
  &= \sum_{X\in\set{\z,+,-}} \abs{g_X}^2 \D\p{s_X} \O
  + \sum_{\mu,j}
  \p{g_\bmu^* g_\mu s_\mu^{(j)} \O s_\mu^{(j)}
    + g_\z^* g_\mu s_\z^{(j)} \O s_\mu^{(j)}
    + g_\bmu^* g_\z s_\mu^{(j)} \O s_\z^{(j)}}
  \notag \\
  &\qquad -\f14 \sum_\mu \mu
  \p{g_\z^*g_\mu - g_\bmu^*g_\z} \sp{S_\mu, \O}_+.
\end{align}
In order to compute the effect of this decoherence on general
collective spin operators, we expand the anti-commutator
\begin{align}
  \sp{S_\mu, S_\mu^\ell \p{\mu S_\z}^m S_\bmu^n}_+
  = S_\mu^{\ell+1} \sp{\p{\mu S_\z}^m+\p{1+\mu S_\z}^m} S_\bmu^n
  - n S_\mu^\ell \p{n-1+2\mu S_\z} \p{\mu S_\z}^m S_\bmu^{n-1}.
  \label{eq:S_mu_acomm}
\end{align}
Recognizing a resemblance between terms in \eqref{eq:S_mu_acomm} and
\eqref{eq:san_z_mu}, we collect terms to simplify
\begin{align}
  \sum_j s_\z^{(j)} S_\mu^\ell \p{\mu S_\z}^m S_\bmu^n s_\mu^{(j)}
  - \f14 \mu \sp{S_\mu, S_\mu^\ell \p{\mu S_\z}^m S_\bmu^n}_+
  = K_{\ell mn;\mu} + L_{\ell mn;\mu}
  \label{eq:dec_z_mu}
\end{align}
and likewise
\begin{align}
  \sum_j s_\mu^{(j)} S_\mu^\ell \p{\mu S_\z}^m S_\bmu^n s_\z^{(j)}
  + \f14 \mu \sp{S_\mu, S_\mu^\ell \p{\mu S_\z}^m S_\bmu^n}_+
  = K_{\ell mn;\mu} + M_{\ell mn;\mu}
  \label{eq:dec_mu_z}
\end{align}
with
\begin{align}
  K_{\ell mn;\mu}
  &\equiv \f14 \mu S_\mu^{\ell+1}
  \sp{\p{1+\mu S_\z}^m-\p{\mu S_\z}^m} S_\bmu^n, \\
  L_{\ell mn;\mu}
  &\equiv -\mu n S_\mu^\ell \sp{S-\ell-\f34\p{n-1}-\f12\mu S_\z}
  \p{\mu S_\z}^m S_\bmu^{n-1}
  - \mu\ell n\p{n-1} S_\mu^{\ell-1}
  \p{S+\mu S_\z} \p{-1+\mu S_\z}^m S_\bmu^{n-2}, \\
  M_{\ell mn;\mu}
  &\equiv \mu n S_\mu^\ell \sp{\p{S+\mu S_\z}\p{-1+\mu S_\z}^m
    - \f12\p{\f12\sp{n-1}+\mu S_\z}\p{\mu S_\z}^m} S_\bmu^{n-1}.
\end{align}
Defining for completion
\begin{align}
  P_{\ell mn;\mu}
  &\equiv \sum_j s_\mu^{(j)} S_\mu^\ell
  \p{\mu S_\z}^m S_\bmu^n s_\mu^{(j)}
  = n S_\mu^{\ell+1} \p{\mu S_\z}^m S_\bmu^{n-1}
  - n\p{n-1} S_\mu^\ell \p{S+\mu S_\z} \p{-1+\mu S_\z}^m S_\bmu^{n-2},
\end{align}
and
\begin{align}
  Q_{\ell mn;\mu}^{(g)}
  \equiv g_\bmu^* g_\mu P_{\ell mn;\mu}
  + \p{g_\z^* g_\mu + g_\bmu^* g_\z}
  K_{\ell mn;\mu}
  + g_\z^* g_\mu L_{\ell mn;\mu}
  + g_\bmu^* g_\z M_{\ell mn;\mu},
  \label{eq:Q_single}
\end{align}
we finally have
\begin{align}
  \D\p{g} \p{S_\mu^\ell \p{\mu S_\z}^m S_\bmu^n}
  = \sum_{X\in\set{\z,+,-}} \abs{g_X}^2
  \D\p{s_X} \p{S_\mu^\ell \p{\mu S_\z}^m S_\bmu^n}
  + Q_{\ell mn;\mu}^{(g)} + \sp{Q_{nm\ell;\mu}^{(g)}}^\dag.
\end{align}
Note that the sum $\ell+m+n$ for operators
$S_\mu^\ell \p{\mu S_\z}^m S_\bmu^n$ does not grow under this type of
decoherence.

%%%%%%%%%%%%%%%%%%%%%%%%%%%%%%%%%%%%%%%%%%%%%%%%%%%%%%%%%%%%%%%%%%%%%%
\section{Sandwich identities for collective-spin decoherence
  calculations}
\label{sec:sandwich_collective}

In analogy with the work in Appendix \ref{sec:sandwich_single}, in
this section we work out sandwich identities necessary for
collective-spin decoherence calculations.  The simplest cases are
\begin{align}
  S_\mu S_\mu^\ell \p{\mu S_\z}^m S_\bmu^n S_\bmu
  &= S_\mu^{\ell+1} \p{\mu S_\z}^m S_\bmu^{n+1},
  \allowdisplaybreaks \\
  S_\mu S_\mu^\ell \p{\mu S_\z}^m S_\bmu^n S_\z
  &= \mu S_\mu^{\ell+1} \p{n+\mu S_\z} \p{\mu S_\z}^m S_\bmu^n,
  \allowdisplaybreaks \\
  S_\z S_\mu^\ell \p{\mu S_\z}^m S_\bmu^n S_\z
  &= S_\mu^\ell \sp{\ell n + \p{\ell+n} \mu S_\z + \p{\mu S_\z}^2}
  \p{\mu S_\z}^m S_\bmu^n.
\end{align}
With a bit more work, we can also find
\begin{align}
  S_\mu^\ell \p{\mu S_\z}^m S_\bmu^n S_\mu
  &= S_\mu^{\ell+1} \p{1+\mu S_\z}^m S_\bmu^n
  - n S_\mu^\ell \p{n-1+2\mu S_\z} \p{\mu S_\z}^m S_\bmu^{n-1},
\end{align}
which implies
\begin{align}
  S_\mu S_\mu^\ell \p{\mu S_\z}^m S_\bmu^n S_\mu
  &= S_\mu^{\ell+2} \p{1+\mu S_\z}^m S_\bmu^n
  - n S_\mu^{\ell+1} \p{n-1+2\mu S_\z} \p{\mu S_\z}^m S_\bmu^{n-1},
  \allowdisplaybreaks \\
  S_\z S_\mu^\ell \p{\mu S_\z}^m S_\bmu^n S_\mu
  &= \mu S_\mu^{\ell+1} \p{\ell+1+\mu S_\z} \p{1+\mu S_\z}^m S_\bmu^n
  \notag \\
  &\qquad - \mu n S_\mu^\ell
  \sp{\ell\p{n-1} + \p{2\ell+n-1}\mu S_\z + 2\p{\mu S_\z}^2}
  \p{\mu S_\z}^m S_\bmu^{n-1}.
\end{align}
Finally, we compute
\begin{align}
  S_\bmu S_\mu^\ell \p{\mu S_\z}^m S_\bmu^n S_\mu
  &= \sp{S_\mu^\ell S_\bmu - \ell S_\mu^{\ell-1} \p{\ell-1+2\mu S_\z}}
  \p{\mu S_\z}^m
  \sp{S_\mu S_\bmu^n - n \p{n-1+2\mu S_\z} S_\bmu^{n-1}} \notag \\
  &= S_\mu^\ell S_\bmu \p{\mu S_\z}^m S_\mu S_\bmu^n \notag \\
  &\qquad - S_\mu^\ell
  \sp{\ell\p{\ell+1} + n\p{n+1}+2\p{\ell+n}\mu S_\z}
  \p{1+\mu S_\z}^m S_\bmu^n \notag \\
  &\qquad + \ell n S_\mu^{\ell-1}
  \sp{\p{\ell-1}\p{n-1}+2\p{\ell+n-2}\mu S_\z + 4\p{\mu S_\z}^2}
  \p{\mu S_\z}^m S_\bmu^{n-1},
\end{align}
where
\begin{align}
  S_\bmu \p{\mu S_\z}^m S_\mu
  = S_\bmu S_\mu \p{1+\mu S_\z}^m
  = \p{S_\mu S_\bmu - 2\mu S_\z} \p{1+\mu S_\z}^m
  = S_\mu \p{2+\mu S_\z}^m S_\bmu - 2\mu S_\z \p{1+\mu S_\z}^m,
\end{align}
so
\begin{align}
  S_\bmu S_\mu^\ell \p{\mu S_\z}^m S_\bmu^n S_\mu
  &= S_\mu^{\ell+1} \p{2+\mu S_\z}^m S_\bmu^{n+1} \notag \\
  &\qquad - S_\mu^\ell
  \sp{\ell\p{\ell+1} + n\p{n+1}+2\p{\ell+n+1}\mu S_\z}
  \p{1+\mu S_\z}^m S_\bmu^n \notag \\
  &\qquad + \ell n S_\mu^{\ell-1}
  \sp{\p{\ell-1}\p{n-1}+2\p{\ell+n-2}\mu S_\z + 4\p{\mu S_\z}^2}
  \p{\mu S_\z}^m S_\bmu^{n-1}.
\end{align}

%%%%%%%%%%%%%%%%%%%%%%%%%%%%%%%%%%%%%%%%%%%%%%%%%%%%%%%%%%%%%%%%%%%%%%
\section{Collective spin decoherence}
\label{sec:decoherence_collective}

In this section we work out the effects of collective decoherence on
general collective spin operators.  For shorthand, we define
\begin{align}
  \D\p{G} \O
  \equiv \D\p{\set{G}} \O
  = G^\dag \O G - \f12\sp{G^\dag G, \O}_+,
\end{align}
where $G$ is a collective spin jump operator.

%%%%%%%%%%%%%%%%%%%%%%%%%%%%%%%%%%%%%%%%%%%%%%%%%%%%%%%%%%%%%%%%%%%%%%
\subsection{Decay-type decoherence and dephasing}
\label{sec:decay_dephasing_collective}

Making use of the results in Appendix \ref{sec:sandwich_collective},
we find that the effects of collective decay-type decoherence on
general collective spin operators are given by
\begin{align}
  \D\p{S_\bmu} \p{S_\mu^\ell \p{\mu S_\z}^m S_\bmu^n}
  &= -S_\mu^{\ell+1} \sp{\p{1+\mu S_\z}^m - \p{\mu S_\z}^m} S_\bmu^{n+1}
  \notag \\
  &\qquad + \f12 S_\mu^\ell \sp{\ell\p{\ell-1} + n\p{n-1}
    + 2\p{\ell+n}\mu S_\z} \p{\mu S_\z}^m S_\bmu^n,
\end{align}
and
\begin{align}
  \D\p{S_\mu} \p{S_\mu^\ell \p{\mu S_\z}^m S_\bmu^n}
  &= S_\mu^{\ell+1} \sp{\p{2+\mu S_\z}^m-\p{1+\mu S_\z}^m} S_\bmu^{n+1}
  \notag \\
  &\qquad - S_\mu^\ell
  \sp{\ell\p{\ell+1} + n\p{n+1}+2\p{\ell+n+1}\mu S_\z}
  \p{1+\mu S_\z}^m S_\bmu^n \notag \\
  &\qquad + \f12 S_\mu^\ell
  \sp{\ell\p{\ell+1} + n\p{n+1}+2\p{\ell+n+2}\mu S_\z}
  \p{\mu S_\z}^m S_\bmu^n \notag \\
  &\qquad + \ell n S_\mu^{\ell-1}
  \sp{\p{\ell-1}\p{n-1}+2\p{\ell+n-2}\mu S_\z + 4\p{\mu S_\z}^2}
  \p{\mu S_\z}^m S_\bmu^{n-1}.
\end{align}
Similarly, the effect of collective dephasing is given by
\begin{align}
  \D\p{S_\z} \p{S_\mu^\ell \p{\mu S_\z}^m S_\bmu^n}
  = -\f12 \p{\ell-n}^2 S_\mu^\ell \p{\mu S_\z}^m S_\bmu^n.
\end{align}

%%%%%%%%%%%%%%%%%%%%%%%%%%%%%%%%%%%%%%%%%%%%%%%%%%%%%%%%%%%%%%%%%%%%%%
\subsection{The general case}
\label{sec:general_collective}

More generally, we consider jump operators of the form
\begin{align}
  G \equiv G_\z S_\z + G_+ S_+ + G_- S_-,
\end{align}
whose decoherence effects are determined by
\begin{align}
  G^\dag \O G
  = \abs{G_\z}^2 S_\z \O S_\z
  + \sum_\mu \p{\abs{G_\mu}^2 S_\bmu \O S_\mu
    + G_\bmu^* G_\mu S_\mu \O S_\mu
    + G_\z^* G_\mu S_\z \O S_\mu
    + G_\bmu^* G_\z S_\mu \O S_\z},
\end{align}
and
\begin{align}
  G^\dag G
  = \abs{G_\z}^2 S_\z^2
  + \sum_\mu \p{\abs{G_\mu}^2 S_\bmu S_\mu
    + G_\z^*G_\mu S_\z S_\mu
    + G_\bmu^* G_\z S_\mu S_\z
    + G_\bmu^* G_\mu S_\mu^2},
\end{align}
which implies
\begin{align}
  \D\p{G} \O
  &= \sum_{X\in\set{\z,+,-}} \abs{G_X}^2 \D\p{S_X} \O
  + \sum_\mu \p{G_\bmu^* G_\mu S_\mu \O S_\mu
    + G_\z^* G_\mu S_\z \O S_\mu
    + G_\bmu^* G_\z S_\mu \O S_\z}
  \notag \\
  &\qquad -\f12 \sum_\mu\p{G_\bmu^* G_\mu \sp{S_\mu^2, \O}_+
    + G_\z^*G_\mu \sp{S_\z S_\mu, \O}_+
    + G_\bmu^* G_\z \sp{S_\mu S_\z, \O}_+}.
\end{align}
In order to compute the effect of this decoherence on general
collective spin operators, we expand the anti-commutators
\begin{align}
  \sp{S_\mu^2, S_\mu^\ell \p{\mu S_\z}^m S_\bmu^n}_+
  &= S_\mu^{\ell+2} \sp{\p{2+\mu S_\z}^m+\p{\mu S_\z}^m} S_\bmu^n
  - 2n S_\mu^{\ell+1} \p{n+2\mu S_\z} \p{1+\mu S_\z}^m S_\bmu^{n-1}
  \notag \\
  &\qquad + n\p{n-1} S_\mu^\ell \sp{\p{n-1}\p{n-2}
    + 2\p{2n-3}\mu S_\z + 4\p{\mu S_\z}^2} \p{\mu S_\z}^m S_\bmu^{n-2},
  \allowdisplaybreaks \\
  \sp{S_\z S_\mu, S_\mu^\ell \p{\mu S_\z}^m S_\bmu^n}_+
  &= \mu S_\mu^{\ell+1} \sp{\p{\ell+1+\mu S_\z}\p{\mu S_\z}^m
    + \p{n+1+\mu S_\z} \p{1+\mu S_\z}^m } S_\bmu^n \notag \\
  &\qquad - \mu n S_\mu^\ell \sp{n \p{n-1}
    + \p{3n-1}\mu S_\z + 2\p{\mu S_\z}^2} \p{\mu S_\z}^m S_\bmu^{n-1},
  \allowdisplaybreaks \\
  \sp{S_\mu S_\z, S_\mu^\ell \p{\mu S_\z}^m S_\bmu^n}_+
  &= \mu S_\mu^{\ell+1} \sp{\p{\ell+\mu S_\z}\p{\mu S_\z}^m
    + \p{n+\mu S_\z} \p{1+\mu S_\z}^m} S_\bmu^n \notag \\
  &\qquad - \mu n S_\mu^\ell \sp{\p{n-1}^2
    + 3\p{n-1}\mu S_\z + 2\p{\mu S_\z}^2} \p{\mu S_\z}^m S_\bmu^{n-1}.
\end{align}
Collecting terms and defining
\begin{align}
  G_{\z,\mu}^{(\pm)}
  &\equiv \f12\p{G_\z^* G_\mu \pm G_\bmu^* G_\z},
  \allowdisplaybreaks \\
  \tilde L_{\ell mn;\mu}^{(G)}
  &\equiv \mu \sp{\p{\ell-n+\f12} G_{\z,\mu}^{(+)}
    + \p{\ell+\f12} G_{\z,\mu}^{(-)}}
  S_\mu^{\ell+1} \p{1+\mu S_\z}^m S_\bmu^n \notag \\
  &\qquad -\mu \sp{\p{\ell-n+\f12} G_{\z,\mu}^{(+)}
    + \p{n+\f12} G_{\z,\mu}^{(-)}}
  S_\mu^{\ell+1} \p{\mu S_\z}^m S_\bmu^n \notag \\
  &\qquad + \mu G_{\z,\mu}^{(-)}
  S_\mu^{\ell+1} \mu S_\z
  \sp{\p{1+\mu S_\z}^m - \p{\mu S_\z}^m} S_\bmu^n,
  \allowdisplaybreaks \\
  \tilde M_{\ell mn;\mu}^{(G)}
  &= -\mu n\p{n-1} \sp{\p{\ell-n+\f12} G_{\z,\mu}^{(+)}
    + \p{\ell-\f12} G_{\z,\mu}^{(-)}}
  S_\mu^\ell \p{\mu S_\z}^m S_\bmu^{n-1} \notag \\
  &\qquad - 2\mu n \sp{\p{\ell-n+\f12} G_{\z,\mu}^{(+)}
    + \p{\ell+\f12n-1} G_{\z,\mu}^{(-)}}
  S_\mu^\ell \p{\mu S_\z}^{m+1} S_\bmu^{n-1} \notag \\
  &\qquad - 2\mu n G_{\z,\mu}^{(-)}
  S_\mu^\ell \p{\mu S_\z}^{m+2} S_\bmu^{n-1},
  \allowdisplaybreaks \\
  \tilde P_{\ell mn;\mu}
  &\equiv -\f12 S_\mu^{\ell+2}
  \sp{\p{2+\mu S_\z}^m - 2\p{1+\mu S_\z}^m + \p{\mu S_\z}^m}
  S_\bmu^n \notag \\
  &\qquad + n S_\mu^{\ell+1} \sp{\p{n+2\mu S_\z} \p{1+\mu S_\z}^m
    - \p{n-1+2\mu S_\z} \p{\mu S_\z}^m}
  S_\bmu^{n-1} \notag \\
  &\qquad -n\p{n-1} S_\mu^\ell
  \sp{\f12\p{n-1}\p{n-2} + \p{2n-3}\mu S_\z + 2\p{\mu S_\z}^2}
  \p{\mu S_\z}^m S_\bmu^{n-2},
  \allowdisplaybreaks \\
  \tilde Q_{\ell mn;\mu}^{(G)}
  &\equiv G_\bmu^* G_\mu \tilde P_{\ell mn;\mu}
  + \tilde L_{\ell mn;\mu}^{(G)}
  + \tilde M_{\ell mn;\mu}^{(G)},
\end{align}
we then have
\begin{align}
  \D\p{G} \p{S_\mu^\ell \p{\mu S_\z}^m S_\bmu^n}
  = \sum_{X\in\set{\z,+,-}} \abs{G_X}^2
  \D\p{S_X} \p{S_\mu^\ell \p{\mu S_\z}^m S_\bmu^n}
  + \tilde Q_{\ell mn;\mu}^{(G)}
  + \sp{\tilde Q_{nm\ell;\mu}^{(G)}}^\dag.
\end{align}
Note that the sum $\ell+m+n$ for operators
$S_\mu^\ell \p{\mu S_\z}^m S_\bmu^n$ grows by one if $G_\mu\ne0$ or
$G_\bmu\ne0$, and does not grow otherwise.

%%%%%%%%%%%%%%%%%%%%%%%%%%%%%%%%%%%%%%%%%%%%%%%%%%%%%%%%%%%%%%%%%%%%%%
\section{Emulating particle loss in a spin model}
\label{sec:particle_loss}

Here we discuss the details of emulating particle loss with $O(1/N)$
error, where $N$ is the initial number of particles in a system that
we wish to describe with a spin model.  Starting with the full algebra
of creation and annihilation operators (whether bosonic or fermionic)
in a system, spin models are typically implemented by identifying a
subalgebra of relevant ``spin'' operators that satisfy appropriate
commutation relations.  Two-state particles on a lattice, for example,
are described by annihilation operators $c_{js}$ indexed by a lattice
site $j\in\mathbb{Z}$ and an internal state index $s\in\set{\up,\dn}$,
enabling the straightforward construction of spin operators
\begin{align}
  \sigma_\x^{(j)} \equiv c_{j,\up}^\dag c_{j,\dn} + \t{h.c.},
  &&
  \sigma_\y^{(j)} \equiv - i c_{j,\up}^\dag c_{j,\dn} + \t{h.c.},
  &&
  \sigma_\z^{(j)}
  \equiv c_{j,\up}^\dag c_{j,\up} - c_{j,\dn}^\dag c_{j,\dn},
  &&
  \1^{(j)} \equiv c_{j,\up}^\dag c_{j,\up} + c_{j,\dn}^\dag c_{j,\dn},
\end{align}
which satisfy the same commutation relations as the standard Pauli
operators.  These spin operators can be more compactly defined in the
form
\begin{align}
  \sigma_\alpha^{(j)} \equiv \sum_{r,s\in\set{\up,\dn}}
  c_{jr}^\dag \bk{r|\sigma_\alpha|s} c_{js},
\end{align}
where $\sigma_\alpha$ for $\alpha\in\set{\x,\y,\z,\1}$ is a Pauli
operator, with $\sigma_\1\equiv\1$; and $\bk{r|\sigma_\alpha|s}$
denotes a matrix element of $\sigma_\alpha$.  This construction
exemplifies how the set of jump operators
$\J_{\t{loss}}^{\t{bare}}\equiv\set{c_{js}}$ that generate particle
loss cannot be constructed from spin operators, which are generally
bilinear in particle creation or annihilation operators.  When working
on the level of a spin model, therefore, we can at best only {\it
  emulate} the effect of particle loss by some indirect means.

To understand the effect of particle loss on collective spin
operators, we first define a single multi-body spin operator
addressing sites $\v j=\set{j_1,j_2,\cdots,j_{\abs{\v j}}}$,
\begin{align}
  \sigma_{\v j\v\alpha} \equiv \prod_{j\in\v j} \sigma_{\alpha_j}^{(j)},
\end{align}
and expand
\begin{align}
  \D\p{\J_{\t{loss}}^{\t{bare}}} \sigma_{\v j\v\alpha}
  &= \sum_{k,s} \p{c_{ks}^\dag \sigma_{\v j\v\alpha} c_{ks}
    - \f12\sp{c_{ks}^\dag c_{ks}, \sigma_{\v j\v\alpha}}_+} \\
  &= \sum_{k\in\v j} \sum_s c_{ks}^\dag \sigma_{\alpha_k}^{(k)} c_{ks}
  \prod_{\substack{j\in\v j\\j\ne k}} \sigma_{\alpha_j}^{(j)}
  + \sum_{k\notin\v j} \sum_s c_{ks}^\dag c_{ks} \sigma_{\v j\v\alpha}
  - \f12 \sum_k \sp{\1^{(k)}, \sigma_{\v j\v\alpha}}_+ \\
  &= \sum_{k\in\v j} \sum_{q,r,s} \bk{q|\sigma_{\alpha_k}|r}
  c_{ks}^\dag c_{kq}^\dag c_{kr} c_{ks}
  \prod_{\substack{j\in\v j\\j\ne k}} \sigma_{\alpha_j}^{(j)}
  - \abs{\v j} \sigma_{\v j\v\alpha}.
\end{align}
In order to have an actual spin model, fermionic statistics or
energetic considerations must forbid multiple occupation of individual
lattice sites.  In that case, the on-site four-point product
$c_{ks}^\dag c_{kq}^\dag c_{kr} c_{ks}=0$ vanishes, and
\begin{align}
  \D\p{\J_{\t{loss}}^{\t{bare}}} \sigma_{\v j\v\alpha}
  = - \abs{\v j} \sigma_{\v j\v\alpha}.
\end{align}
Up to $O(1/N)$ corrections, a collective spin operator $\S_{\v m}$
essentially consists of $\abs{\v m}$-body operators of the form
$\sigma_{\v j\v\alpha}$ with $\abs{\v j}=\abs{\v m}$, which implies
that the dissipator $\D_{\t{loss}}$ defined by
$\D_{\t{loss}}\S_{\v m}=-\abs{\v m}\S_{\v m}$ describes particle loss
with $O(1/N)$ error.  We note that the dissipator $\D_{\t{loss}}$ is
essentially the depolarizing channel,
i.e.~$\D_{\t{loss}}=\D\p{\J_{\t{loss}}}$ for
$\J_{\t{loss}} = \set{s_\alpha^{(j)}}$ with $\alpha\in\set{\x,\y,\z}$
and $j\in\set{1,2,\cdots,N}$.  A direct implementation of
$\D_{\t{loss}}$ with $\D_{\t{loss}}\S_{\v m}=-\abs{\v m}\S_{\v m}$,
however, is much more efficient than evaluating the depolarizing
channel $\D\p{\J_{\t{loss}}}$ with the ingredients in Appendices
\ref{sec:sandwich_single} and \ref{sec:decoherence_single}.

%%%%%%%%%%%%%%%%%%%%%%%%%%%%%%%%%%%%%%%%%%%%%%%%%%%%%%%%%%%%%%%%%%%%%%
\section{Initial conditions}
\label{sec:initial_conditions}

Here we compute the expectation values of collective spin operators
with respect to spin-polarized (also Gaussian, or spin-coherent)
states.  These states are parameterized by polar and azimuthal angles
$\theta\in[0,\pi)$, $\phi\in[0,2\pi)$, and lie within the Dicke
manifold spanned by states
$\ket{k}\propto S_+^{S+k}\ket{\dn}^{\otimes N}$ with $S\equiv N/2$ and
$S_\z\ket{k}=k\ket{k}$:
\begin{align}
  \ket{\theta,\phi}
  \equiv \sp{\cos\p{\theta/2} e^{-i\phi/2} \ket\up
    + \sin\p{\theta/2} e^{i\phi/2} \ket\dn}^{\otimes N}
  = \sum_{k=-S}^S { N \choose S+k }^{1/2}
  \cos\p{\theta/2}^{S+k} \sin\p{\theta/2}^{S-k} e^{-ik\phi} \ket{k}.
\end{align}
We can likewise expand, within the Dicke manifold,
\begin{align}
  S_\z = \sum_{k=-S}^S k \op{k},
  &&
  S_\mu = \sum_{k=-S+\delta_{\mu,-1}}^{S-\delta_{\mu,1}}
  g_\mu\p{k} \op{k+\mu}{k}
  = \sum_{k=-S+\delta_{\bmu,-1}}^{S-\delta_{\bmu,1}}
  g_\bmu\p{k} \op{k}{k+\bmu},
\end{align}
where $\bmu\equiv-\mu\in\set{+1,-1}$ and
\begin{align}
  g_\mu\p{k} \equiv \sqrt{\p{S-\mu k}\p{S+\mu k+1}},
\end{align}
which implies
\begin{align}
  S_\mu^\ell \p{\mu S_\z^m} S_\bmu^n
  &= \sum_{k=-S+\delta_{\mu,-1}\max\set{\ell,n}}
  ^{S-\delta_{\mu,1}\max\set{\ell,n}} \p{\mu k}^m
  \sp{\prod_{p=0}^{\ell-1} g_\mu\p{k+\mu p}}
  \sp{\prod_{q=0}^{n-1} g_\mu\p{k+\mu q}}
  \op{k+\mu\ell}{k+\mu n} \\
  &= \sum_{\mu k=-\mu S-\delta_{\mu,-1}\max\set{\ell,n}}
  ^{\mu S-\delta_{\mu,1}\max\set{\ell,n}} \p{\mu k}^m
  \f{\p{S-\mu k}!}{\p{S+\mu k}!}
  \sp{\f{\p{S+\mu k+\ell}!}{\p{S-\mu k-\ell}!}
    \f{\p{S+\mu k+n}!}{\p{S-\mu k-n}!}}^{1/2}
  \op{k+\mu\ell}{k+\mu n} \\
  &= \sum_{k=-S}^{S-\max\set{\ell,n}} k^m
  \f{\p{S-k}!}{\p{S+k}!}
  \sp{\f{\p{S+k+\ell}!}{\p{S-k-\ell}!}
    \f{\p{S+k+n}!}{\p{S-k-n}!}}^{1/2}
  \op{\mu\p{k+\ell}}{\mu\p{k+n}}.
\end{align}
This expansion allows us to compute the expectation value
\begin{align}
  \bk{\theta,\phi|S_\mu^\ell \p{\mu S_\z^m} S_\bmu^n|\theta,\phi}
  &= e^{i\phi \mu\p{\ell-n}} N! \sum_{k=-S}^{S-\max\set{\ell,n}}
  \f{k^m \p{S-k}! f_{\mu\ell n}\p{k,\theta}}
  {\p{S+k}! \p{S-k-\ell}! \p{S-k-n}!} \\
  &= e^{i\phi \mu\p{\ell-n}} \p{-1}^m N! \sum_{k=0}^{N-\max\set{\ell,n}}
  \f{\p{S-k}^m \p{N-k}! \tilde f_{\mu\ell n}\p{k,\theta}}
  {k! \p{N-k-\ell}! \p{N-k-n}!}
\end{align}
where
\begin{align}
  f_{\mu\ell n}\p{k,\theta}
  \equiv \cos\p{\theta/2}^{N+\mu\p{2k+\ell+n}}
  \sin\p{\theta/2}^{N-\mu\p{2k+\ell+n}},
\end{align}
\begin{align}
  \tilde f_{\mu\ell n}\p{k,\theta}
  \equiv f_{\mu\ell n}\p{k-S,\theta}
  = \cos\p{\theta/2}^{2N\delta_{\mu,-1}+\mu\p{2k+\ell+n}}
  \sin\p{\theta/2}^{2N\delta_{\mu,1}-\mu\p{2k+\ell+n}}.
\end{align}
Defining the states
\begin{align}
  \ket{+\Z} \equiv \ket{0,0} = \ket\up^{\otimes N}, &&
  \ket{-\Z} \equiv \ket{\pi,0} = \ket\dn^{\otimes N}, &&
  \ket\X \equiv \ket{\pi/2,0}
  = \p{\f{\ket\up+\ket\dn}{\sqrt2}}^{\otimes N},
\end{align}
some particular expectation values of interest are
\begin{align}
  \bk{\nu\Z|S_\mu^\ell \p{\mu S_\z}^m S_\bmu^n|\nu\Z}
  = \delta_{\ell n} \times
  \begin{cases}
    \p{S-n}^m \f{N! n!}{\p{N-n}!} & \mu = \nu, \\
    \delta_{n,0} \p{-S}^m & \mu \ne \nu,
  \end{cases},
\end{align}
and
\begin{align}
  \bk{\X|S_\mu^\ell \p{\mu S_\z}^m S_\bmu^n|\X}
  = \p{-1}^m \f{N!}{2^N} \sum_{k=0}^{N-\max\set{\ell,n}}
  \f{\p{S-k}^m \p{N-k}!}{k!\p{N-k-\ell}!\p{N-k-n}!}.
\end{align}

%%%%%%%%%%%%%%%%%%%%%%%%%%%%%%%%%%%%%%%%%%%%%%%%%%%%%%%%%%%%%%%%%%%%%%
\section{Computing correlators with the truncated short-time (TST)
  expansion}
\label{sec:tutorial}

Here we provide a pedagogical tutorial for computing correlators using
the truncated short-time TST expansion.  For concreteness, we
nominally consider $N$ spins evolving under the one-axis twisting
(OAT) Hamiltonian
\begin{align}
  H_{\t{OAT}} = \chi S_\z^2,
\end{align}
additionally subject to spontaneous single-spin decay at rate
$\gamma_-$, with jump operators $\J_-=\set{s_-^{(j)}:j=1,2,\cdots,N}$.
The equation of motion for a Heisenberg operator
$\p{S_+^\ell S_\z^m S_-^n}\p{t}$ is
\begin{align}
  \f{d}{dt} \Bk{S_+^\ell S_\z^m S_-^n}
  = i\chi\Bk{\sp{S_\z^2, S_+^\ell S_\z^m S_-^n}_-}
  + \gamma_- \Bk{\D\p{\J_-}\p{S_+^\ell S_\z^m S_-^n}},
  \label{eq:tutorial_EOM}
\end{align}
where we have suppressed the explicit time dependence of operators for
brevity.  Using the results in appendices \ref{sec:general_product}
and \ref{sec:decay_single} respectively to evaluate the commutator
$\sp{S_\z^2, S_+^\ell S_\z^m S_-^n}_-$ and dissipator
$\D\p{\J_-} \p{S_+^\ell S_\z^m S_-^n}$ in \eqref{eq:tutorial_EOM}, we
can expand
\begin{multline}
  \f{d}{dt} \bk{S_+^\ell S_\z^m S_-^n} \\
  = i\chi \bk{\p{\ell-n} S_+^\ell \p{\ell+n+2S_\z} S_\z^m S_-^n}
  + \gamma_- \Bk{S_+^\ell \sp{\p{S+S_\z}\p{-1+S_\z}^m
      - \p{S+\f{\ell+n}{2}+S_\z} S_\z^m} S_-^n}.
  \label{eq:tutorial_EOM_expanded}
\end{multline}
In practice, we do not want to keep track of such an expansion by
hand, especially in the case of e.g.~the two-axis twisting (TAT) and
twist-and-turn (TNT) models with more general types of decoherence,
for which the analogue of \eqref{eq:tutorial_EOM_expanded} may take
several lines just to write out in full.  Defining the operators
$\S_{\v m}\equiv S_+^{m_+} S_\z^{m_\z} S_-^{m_-}$ with
$\v m\equiv\p{m_+,m_\z,m_-}$ for shorthand, we note that the vector
space spanned by $\set{\S_{\v m}}$ is closed under time evolution.  We
therefore expand
\begin{align}
  \f{d}{dt} \bk{\S_{\v n}}
  = \bk{T \S_{\v n}}
  = \sum_{\v m} \bk{\S_{\v m}} T_{\v m\v n},
  \label{eq:tutorial_EOM_general}
\end{align}
where $T$ is a superoperator that generates time evolution for
Heisenberg operators.  In the present example, the matrix elements
$T_{\v m\v n}\in\C$ of $T$ are defined by
\eqref{eq:tutorial_EOM_expanded} and \eqref{eq:tutorial_EOM_general}.
For any Hamiltonian $H$ with decoherence characterized by sets of jump
operators $\J$ and decoherence rates $\gamma_\J$, the matrix elements
$T_{\v m\v n}$ are more generally defined by
\begin{align}
  T \S_{\v n}
  = i\sp{H,\S_{\v n}}_-
  + \sum_\J \gamma_\J \D\p{\J} \S_{\v n}
  = \sum_{\v m} \S_{\v m} T_{\v m\v n}.
\end{align}
The results in Appendices \ref{sec:general_product},
\ref{sec:decoherence_single}, and \ref{sec:decoherence_collective} can
be used to write model-agnostic codes that compute matrix elements
$T_{\v m\v n}$, taking a particular Hamiltonian $H$ and decoherence
processes $\set{\p{\J,\gamma_\J}}$ as inputs.

In order to compute a quantity such as spin squeezing, we need to
compute correlators of the form $\bk{\S_{\v n}\p{t}}$, where for
clarity we will re-introduce the explicit time dependence of
Heisenberg operators $\S_{\v n}\p{t}$.  The order-$M$ truncated
short-time (TST) expansion takes
\begin{align}
  \bk{\S_{\v n}\p{t}}
  = \bk{e^{tT}\S_{\v n}\p{0}}
  = \sum_{k\ge0} \f{t^k}{k!} \bk{T^k\S_{\v n}\p{0}}
  = \sum_{k\ge0} \f{t^k}{k!}
  \sum_{\v m} \bk{\S_{\v m}\p{0}} T^k_{\v m\v n}
  \to \sum_{k=0}^M \f{t^k}{k!}
  \sum_{\v m} \bk{\S_{\v m}\p{0}} T^k_{\v m\v n},
  \label{eq:tutorial_TST}
\end{align}
where $T^k_{\v m\v n}$ are matrix elements of the $k$-th time
derivative operator $T^k$, given by
\begin{align}
  T^0_{\v m\v n} \equiv
  \begin{cases}
    1 & \v m = \v n, \\
    0 & \t{otherwise}
  \end{cases},
  &&
  T^1_{\v m\v n} \equiv T_{\v m\v n},
  &&
  T^{k>1}_{\v m\v n}
  \equiv \sum_{\v p_1,\v p_2,\cdots,\v p_{k-1}}
  T_{\v m\v p_{k-1}} \cdots T_{\v p_3\v p_2}
  T_{\v p_2\v p_1} T_{\v p_1\v n}.
\end{align}
Matrix elements $T^k_{\v m\v n}$ and initial-time expectation values
$\bk{\S_{\v m}\p{0}}$ are thus computed as needed for any particular
correlator $\bk{\S_{\v n}\p{t}}$ of interest, and combined according
to \eqref{eq:tutorial_TST}.  Note that initial-time expectation values
$\bk{\S_{\v m}\p{0}}$ are an {\it input} to the TST expansion, and
need to be computed separately for any initial state of interest;
expectation values with respect to spin-polarized (Gaussian) states
are provided in Appendix \ref{sec:initial_conditions}.  In practice,
we further collect terms in \eqref{eq:tutorial_TST} to write
\begin{align}
  \bk{\S_{\v n}\p{t}} \to \sum_{k=0}^M c_{\v n k} t^k,
  &&
  c_{\v n k}
  \equiv \f1{k!} \sum_{\v m} \bk{\S_{\v m}\p{0}} T^k_{\v m\v n},
  \label{eq:tutorial_corr}
\end{align}
where $c_{\v n k}$ are time-independent coefficients for the expansion
of $\bk{\S_{\v n}\p{t}}$.  After computing the coefficients
$c_{\v n k}$, there is only negligible computational overhead to
compute the correlator $\bk{\S_{\v n}\p{t}}$ for any time $t$.

%%%%%%%%%%%%%%%%%%%%%%%%%%%%%%%%%%%%%%%%%%%%%%%%%%%%%%%%%%%%%%%%%%%%%%
\section{Analytical results for the one-axis twisting model}
\label{sec:OAT}

The one-axis twisting (OAT) Hamiltonian for $N$ spin-1/2 particles
takes the form
\begin{align}
  H_{\t{OAT}}
  = \chi S_\z^2
  = \f12 \chi \sum_{j<k} \sigma_\z^{(j)} \sigma_\z^{(k)} + \f14 N \chi,
\end{align}
where $\sigma_\z^{(j)}$ represents a Pauli-$z$ operator acting on spin
$j$.  This model is a special case of the zero-field Ising Hamiltonian
previously solved in Ref.~[\citenum{foss-feig2013nonequilibrium}] via
exact, analytical treatment of the quantum trajectory Monte Carlo
method for computing expectation values.  The solution therein
accounts for coherent evolution in addition to decoherence via
uncorrelated single-spin decay, excitation, and dephasing respectively
at rates $\gamma_-$, $\gamma_+$, and $\gamma_\z$ (denoted by
$\Gamma_{\t{ud}}$, $\Gamma_{\t{du}}$, and $\Gamma_{\t{el}}$ in
Ref.~[\citenum{foss-feig2013nonequilibrium}]).  Letting $S\equiv N/2$
and $\mu,\nu\in\set{+1,-1}$, we adapt expectation values computed in
Ref.~[\citenum{foss-feig2013nonequilibrium}] for the initial state
$\ket\X\propto\p{\ket\up+\ket\dn}^{\otimes N}$ with
$S_\x\ket\X=S\ket\X$ evolving under $H_{\t{OAT}}$, finding
\begin{align}
  \bk{S_+\p{t}}
  &= S e^{-\kappa t} \Phi\p{\chi,t}^{N-1},
  \label{eq:S+_OAT} \\
  \bk{\p{S_\mu S_\z}\p{t}}
  &= -\f{\mu}{2}\bk{S_\mu\p{t}}
  + S \p{S-\f12} e^{-\kappa t}
  \Psi\p{\mu\chi,t} \Phi\p{\chi,t}^{N-2}, \\
  \bk{\p{S_\mu S_\nu}\p{t}}
  &= \p{1-\delta_{\mu\nu}} \p{S + \mu\bk{S_\z\p{t}}}
  + S \p{S-\f12} e^{-2\kappa t} \Phi\p{\sp{\mu+\nu}\chi,t}^{N-2},
  \label{eq:SS+-_OAT}
\end{align}
where
\begin{align}
  \Phi\p{X,t}
  \equiv e^{-\lambda t}
  \sp{\cos\p{\omega_X t} + \f{\lambda}{\omega_X} \sin\p{\omega_X t}},
  &&
  \Psi\p{X,t}
  \equiv e^{-\lambda t} \p{\f{\Delta+iX}{\omega_X}} \sin\p{\omega_X t},
\end{align}
for
\begin{align}
  \kappa \equiv \f12 \p{\gamma_+ + \gamma_- + \gamma_\z},
  &&
  \lambda \equiv \f12 \p{\gamma_+ + \gamma_-},
  &&
  \Delta \equiv \gamma_+ - \gamma_-,
  &&
  \omega_X \equiv \sqrt{X^2-\lambda^2-iX\Delta}.
\end{align}
In order to compute spin squeezing as measured by the Ramsey squeezing
parameter $\xi^2$ defined in \eqref{eq:squeezing}, we additionally
need analytical expressions for $\bk{S_\z\p{t}}$ and
$\bk{S_\z^2\p{t}}$.  As these operators commute with both the OAT
Hamiltonian and the single-spin operators $\sigma_\z^{(j)}$, their
evolution is governed entirely by decay-type decoherence (see Appendix
\ref{sec:decay_single}), which means
\begin{align}
  \f{d}{dt} \bk{S_\z\p{t}}
  &= S\p{\gamma_+-\gamma_-} - \p{\gamma_++\gamma_-} \bk{S_\z\p{t}},
  \\
  \f{d}{dt} \bk{S_\z^2\p{t}}
  &= S\p{\gamma_++\gamma_-}
  + 2\p{S-\f12}\p{\gamma_+-\gamma_-} \bk{S_\z\p{t}}
  - 2 \p{\gamma_++\gamma_-} \bk{S_\z^2\p{t}}.
\end{align}
The initial conditions $\bk{S_\z\p{0}}=0$ and $\bk{S_\z^2\p{0}}=S/2$
then imply
\begin{align}
  \bk{S_\z\p{t}} = S\p{\f{\gamma_+-\gamma_-}{\gamma_++\gamma_-}}
  \p{1-e^{-\p{\gamma_+ + \gamma_-} t}},
  &&
  \bk{S_\z^2\p{t}}
  = \f12 S + S \p{S-\f12} \p{\f{\bk{S_\z\p{t}}}{S}}^2.
  \label{eq:Sz_OAT}
\end{align}
With appropriate assumptions about the relevant sources of
decoherence, the expectation values in
\eqref{eq:S+_OAT}--\eqref{eq:SS+-_OAT} and \eqref{eq:Sz_OAT} are
sufficient to compute the spin squeezing parameter $\xi^2$ in
\eqref{eq:squeezing} at any time throughout evolution of the initial
state $\ket\X$ under $H_{\t{OAT}}$.

%%%%%%%%%%%%%%%%%%%%%%%%%%%%%%%%%%%%%%%%%%%%%%%%%%%%%%%%%%%%%%%%%%%%%%
\section{Diagnosing breakdown of the TST expansion}
\label{sec:breakdown}

\begin{figure}
  \centering
  \includegraphics{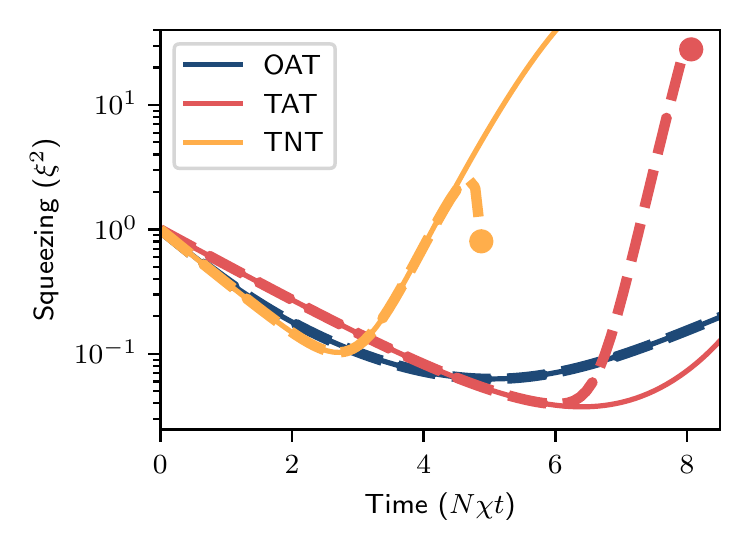}
  \caption{Spin squeezing throughout decoherence-free evolution of
    $N=100$ spins initially in the state $\ket\X$, computed using both
    exact methods (solid lines) and the TST expansion in
    Eq.~\eqref{eq:TST} with $M=35$ (dashed lines).  Solid circles mark
    the times at which the TST expansion gives an unphysical result
    with $\xi^2<0$.}
  \label{fig:unitary_small}
\end{figure}

\begin{figure}
  \centering
  \includegraphics{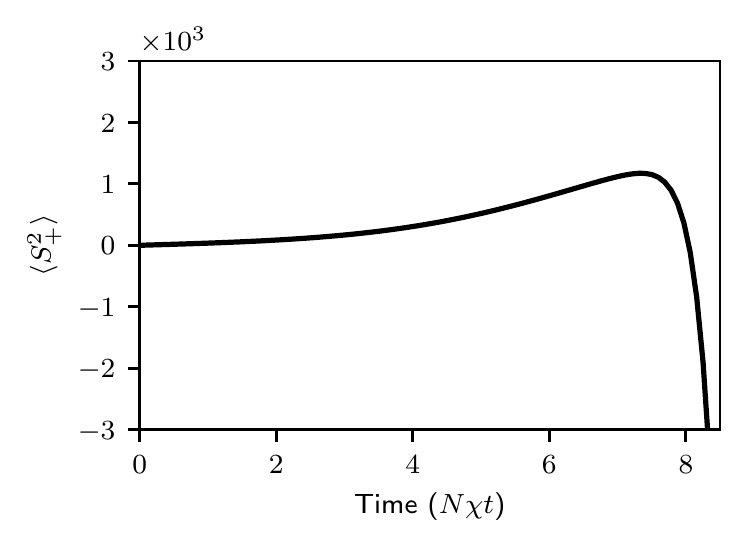}
  \caption{A collective spin correlator in the TAT model with $N=100$
    spins and no decoherence, computed using the TST expansion with
    $M=35$.  The divergence of correlators of this sort can be used to
    diagnose the breakdown of the TST expansion.}
  \label{fig:correlator_example}
\end{figure}

In Figure \ref{fig:benchmarking} of the main text, the TST expansion
provided nearly exact results for squeezing until a sudden departure
that quickly resulted in an unphysical squeezing parameter, $\xi^2<0$.
In general, however, there is no fundamental relationship between the
breakdown of the TST expansion and the conditions for a physical
squeezing parameter $\xi^2$.  A proper diagnosis of breakdown
therefore requires inspection of the correlators $\bk{\S_{\v n}\p{t}}$
used to compute the squeezing parameter $\xi^2$, which upon breakdown
will rapidly take unphysical values with
$\abs{\bk{\S_{\v n}\p{t}}}\gtrsim S^{\abs{\v n}}$.  As an example,
Figure \ref{fig:unitary_small} shows the squeezing parameter $\xi^2$
throughout decoherence-free evolution of $N=100$ spins initially in
the state $\ket\X$.  In this example, the squeezing computed by the
TST expansion for the TAT model diverges from the exact answer without
an immediate and obvious signature of breakdown.  Nonetheless,
breakdown can still be diagnosed by inspection of individual
correlators, as shown in Figure \ref{fig:correlator_example}, where we
plot $\t{Im}\bk{S_+^2}$ as a function of time for $N=100$ spins
evolving under the TAT without decoherence.  Figure
\ref{fig:correlator_example} shows that breakdown clearly occurs
around $N\chi t\lesssim7$, when the correlator $\bk{S_+^2}$ begins to
diverge to values $\gtrsim S^2=\p{N/2}^2=2500$ in magnitude.  A joint
inspection of figures \ref{fig:unitary_small} and
\ref{fig:correlator_example} suffice to trace the anomalous behavior
of $\xi^2$ from $N\chi t\approx 7$ back to $N\chi t\approx 6$, when it
first took a sudden turn before becoming unphysical at
$N\chi t\approx 8$.

%%%%%%%%%%%%%%%%%%%%%%%%%%%%%%%%%%%%%%%%%%%%%%%%%%%%%%%%%%%%%%%%%%%%%%
\section{Spin squeezing with strong decoherence}
\label{sec:strong_dec}

\begin{figure}
  \centering
  \includegraphics{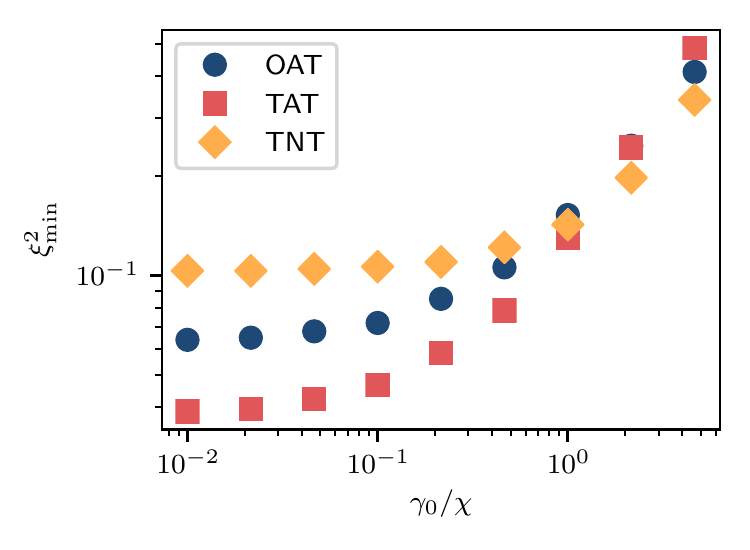}
  \caption{Optimal spin squeezing of $N=100$ spins undergoing
    spontaneous decay, excitation, and dephasing at rates
    $\gamma_-=\gamma_+=\gamma_\z=\gamma_0$, computed using quantum
    trajectory simulations with $10^3$ trajectories per data point.}
  \label{fig:sqz_dec_scaling}
\end{figure}

Here we provide supplementary evidence of our finding in Section
\ref{sec:squeezing} that the TNT model can produce more squeezing than
the OAT or TAT models in the presence of strong decoherence.  To this
end, Figure \ref{fig:sqz_dec_scaling} shows the minimal squeezing
parameter $\xi^2_{\t{min}}$ achievable with $N=100$ spins through the
OAT, TAT, and TNT models as a function of the rate $\gamma_0$ at which
individual spins undergo spontaneous decay, excitation, and dephasing.
These results were computed with quantum trajectory simulations, with
$10^3$ trajectories per data point.  While the OAT and TAT models
produce more squeezing than the TNT model with weak decoherence, this
squeezing falls off faster with an increasing decoherence rate
$\gamma_0$.  The relative robustness of TNT is in part a consequence
of the fact that TNT initially generates squeezing at a faster rate
than OAT or TAT, thereby allowing it to produce more squeezing before
the degrading effects of decoherence kick in.

%%%%%%%%%%%%%%%%%%%%%%%%%%%%%%%%%%%%%%%%%%%%%%%%%%%%%%%%%%%%%%%%%%%%%%
\section{Heisenberg operators in open quantum systems}
\label{sec:noise}

Here we explain the origin and character of the mean-zero ``noise''
operators $\E_\O\p{t}$ that appear in the expansion of a Heisenberg
operator $\O\p{t}=\sum_{\v m}\O_{\v m}\p{t}\S_{\v m}+\E_\O\p{t}$ with
time-dependent coefficients $\O_{\v m}\p{t}$ for time-independent
Schr\"odinger operators $\S_{\v m}$.  Our discussion should clarify
why noise operators play no role in our calculation of expectation
values of the form $\bk{\O\p{t}}$ and $\bk{\O\Q\p{t}}$, despite the
fact that noise operators generally {\it do} need to be considered in
the calculation of more general multi-time correlators in open quantum
systems\cite{blocher2019quantum}.

In any closed quantum system with initial state $\rho$ and propagator
$U\p{t}$, such that the state at time $t$ is
$\rho\p{t}\equiv U\p{t}\rho U^\dag\p{t}$, time-dependent Heisenberg
operators $\O\p{t}$ are uniquely defined from time-independent
Schr\"odinger operators $\O$ by
\begin{align}
  \bk{\O\p{t}} \equiv \tr\sp{\rho\p{t}\O} = \tr\sp{\rho\O\p{t}}.
  \label{eq:heisenberg_condition}
\end{align}
Enforcing \eqref{eq:heisenberg_condition} for {\it arbitrary} initial
states $\rho$ forces $\O\p{t}=U^\dag\p{t}\O U\p{t}$.  In an open
quantum system, however, the definition of a Heisenberg operator is
not so straightforward.  Open systems can often be understood as
subsystems of a larger closed system.  Consider therefore an open
system $S$ with environment $E$, a joint initial state $\rho_{SE}$,
and propagator $U_{SE}\p{t}$.  The reduced state $\rho_S\p{t}$ of $S$
at time $t$ is
\begin{align}
  \rho_S\p{t}
  \equiv \tr_E\sp{\rho_{SE}\p{t}}
  = \tr_E\sp{U_{SE}\p{t} \rho_{SE} U_{SE}^\dag\p{t}}
  \equiv \U_{\ul S}\p{t} \rho_S,
  \label{eq:state_S}
\end{align}
where $\rho_S\equiv\rho_S\p{0}$ is a time-independent state of $S$ in
the Heisenberg picture, $\ul S$ denotes the space of operators on $S$,
and the quantum channel $\U_{\ul S}\p{t}$ has the
decomposition\cite{rivas2012time}
\begin{align}
  \U_{\ul S}\p{t} \rho_S
  = \sum_j \U_S^{(j)}\p{t} \rho_S \U_S^{(j)\dag}\p{t}
\end{align}
with ordinary operators $\U_S^{(j)}\p{t}$ on $S$.  We can therefore
expand
\begin{align}
  \bk{\O_S\p{t}}
  = \tr\sp{\rho_S\p{t}\O_S}
  = \tr\sp{\U_{\ul S}\p{t} \rho_S \O_S}
  = \tr\sp{\rho_S \U_{\ul S}^\dag\p{t} \O_S}
  = \tr\sp{\rho_S \ul{\O_S}\p{t}}
  = \bk{\ul{\O_S}\p{t}},
  \label{eq:heisenberg_correlator}
\end{align}
where $\U_{\ul S}^\dag\p{t}$ is the adjoint map of $\U_{\ul S}\p{t}$
(with respect to a trace inner product between operators on $S$), and
we define the time-dependent operator
\begin{align}
  \ul{\O_S}\p{t}
  \equiv \U_{\ul S}^\dag\p{t} \O_S
  = \sum_j \U_S^{(j)\dag}\p{t} \O_S \U_S^{(j)}\p{t}.
  \label{eq:heisenberg_wrong}
\end{align}
We thus find that substituting $\ul{\O_S}\p{t}$ in place of
$\O_S\p{t}$ suffices for the calculation of correlators
$\bk{\O_S\p{t}}$, thereby accounting for the validity of the equation
of motion in \eqref{eq:EOM}.  As we show below, this substitution also
suffices for the calculation of two-time correlators of the form
$\bk{\O_S\Q_S\p{t}}$ when the environment $E$ is Markovian.

The problem with {\it defining} Heisenberg operators $\O_S\p{t}$ by
$\ul{\O_S}\p{t}$ only becomes evident when considering products of
Heisenberg operators.  One would like for the product of two
Heisenberg operators $\O_S\p{t}$ and $\Q_S\p{t}$ to satisfy
$\O_S\p{t}\Q_S\p{t}=\p{\O_S\Q_S}\p{t}$.  This intuition can be
formalized by observing that
\begin{align}
  \bk{\O_S\p{t}}
  = \tr\sp{\rho_{SE}\p{t} \p{\O_S\otimes\1_E}}
  = \tr\sp{\rho_{SE} \p{\O_S\otimes\1_E}\p{t}}
  = \bk{\p{\O_S\otimes\1_E}\p{t}},
  \label{eq:heisenberg_expansion}
\end{align}
where $\1_E$ is the identity operator on $E$, expectation values of
Heisenberg operators on system $A\in\set{S,E,SE}$ are taken with
respect to the state $\rho_A$, and
\begin{align}
  \p{\O_S\otimes\1_E}\p{t}
  \equiv U_{SE}^\dag\p{t} \p{\O_S\otimes\1_E} U_{SE}\p{t}.
  \label{eq:heisenberg_extension}
\end{align}
By expanding Heisenberg operators similarly to
\eqref{eq:heisenberg_expansion} and \eqref{eq:heisenberg_extension},
we then find
\begin{align}
  \bk{\O_S\p{t}\Q_S\p{t}}
  = \bk{\p{\O_S\otimes\1_E}\p{t} \p{\Q_S\otimes\1_E}\p{t}}
  = \bk{\p{\O_S\Q_S\otimes\1_E}\p{t}}
  = \bk{\p{\O_S\Q_S}\p{t}}.
\end{align}
The expression in \eqref{eq:heisenberg_wrong}, however, makes it clear
that generally
$\ul{\O_S}\p{t} \ul{\Q_S}\p{t}\ne\p{\ul{\O_S\Q_S}}\p{t}$.  To correct
for this discrepancy, we define
\begin{align}
  \O_S\p{t} \equiv \ul{\O_S}\p{t} + \E_{\O_S}\p{t}
  \label{eq:noise_def}
\end{align}
in terms of new ``noise'' operators $\E_{\O_S}\p{t}$ that are
essentially defined to enforce the consistency of operator products
such as $\O_S\p{t}\Q_S\p{t}=\p{\O_S\Q_S}\p{t}$. Self-consistency
forces noise operators to be mean-zero, as
\begin{align}
  \bk{\E_{\O_S}\p{t}} = \bk{\O_S\p{t}} - \bk{\ul{\O_S}\p{t}} = 0.
\end{align}
Furthermore, if the environment $E$ is Markovian, then noise operators
are also uncorrelated with initial-time observables,
i.e.~$\bk{\O_S\E_{\Q_S}\p{t}}=0$, which means that noise operators can
be neglected in the calculation of two-time correlators of the form
$\bk{\O_S\Q_S\p{t}}$.  To see why, we observe that a Markovian
environment is essentially defined to satisfy
\begin{align}
  \rho_{SE}\p{t} = U_{SE}\p{t} \rho_{SE} U_{SE}^\dag\p{t}
  \approx \rho_S\p{t}\otimes\rho_E
  = \U_{\ul S}\p{t} \rho_S\otimes\rho_E,
  \label{eq:markov}
\end{align}
with $\rho_E$ a time-independent steady state of the environment.  If
we enforce \eqref{eq:markov} for all states $\rho_S$, e.g.~the
maximally mixed state $\rho_S^{(1)}\propto\1_S$ and
$\rho_S^{(2)}\equiv\rho_S^{(1)}+\O_S$ with $\O_S$ any traceless
operator on $S$ with operator norm $\norm{\O_S}\le1/\tr\1_S$
(i.e.~such that $\rho_S^{(2)}$ remains positive semi-definite, or a
valid quantum state), then by linearity we find that
\begin{align}
  U_{SE}\p{t} \p{\1_S\otimes\rho_E} U_{SE}^\dag\p{t}
  \approx \U_{\ul S}\p{t} \1_S\otimes\rho_E,
  &&
  U_{SE}\p{t} \p{\O_S\otimes\rho_E} U_{SE}^\dag\p{t}
  \approx \U_{\ul S}\p{t} \O_S\otimes\rho_E,
\end{align}
which implies that the Markov approximation \eqref{eq:markov} holds
even if we replace $\rho_S$ by any operator on $S$, and in particular
\begin{align}
  U_{SE}\p{t} \rho_{SE} \p{\O_S\otimes\1_E} U_{SE}^\dag\p{t}
  = U_{SE}\p{t} \p{\rho_S\O_S\otimes\rho_E} U_{SE}^\dag\p{t}
  \approx \U_{\ul S}\p{t}\p{\rho_S\O_S}\otimes\rho_E.
  \label{eq:makov_op}
\end{align}
We can therefore expand
\begin{align}
  \bk{\O_S\Q_S\p{t}}
  &= \tr\sp{\rho_{SE} \p{\O_S\otimes\1_E}
    U_{SE}^\dag\p{t} \p{\Q_S\otimes\1_E} U_{SE}\p{t}} \\
  &= \tr\sp{U_{SE}\p{t} \rho_{SE} \p{\O_S\otimes\1_E}
    U_{SE}^\dag\p{t} \p{\Q_S\otimes\1_E}},
\end{align}
and invoke the Markov approximation in \eqref{eq:makov_op} to find
that
\begin{align}
  \bk{\O_S\Q_S\p{t}}
  \approx \tr\sp{\U_{\ul S}\p{t}\p{\rho_S\O_S} \Q_S}
  = \tr\sp{\rho_S \O_S \U_{\ul S}^\dag\p{t} \Q_S}
  = \bk{\O_S\ul{\Q_S}\p{t}},
\end{align}
which implies
\begin{align}
  \bk{\O_S\E_{\Q_S}\p{t}}
  = \bk{\O_S\Q_S\p{t}} - \bk{\O_S\ul{\Q_S}\p{t}}
  \approx 0.
\end{align}
Noise operators thus play no role in the calculation of correlators
such as $C\p{t}$ in \eqref{eq:two_time}.  In contrast, noise operators
generally {\it do} play a role in the calculation of multi-time
correlators of the form
$\bk{\prod_j\O_S^{(j)}\p{t_j}}$\cite{blocher2019quantum}.
Furthermore, these calculations generally require additional
assumptions about the environment.  To keep our discussion simple and
general, we therefore exclude the effects of noise terms in Section
\ref{sec:multi_time}.

%%%%%%%%%%%%%%%%%%%%%%%%%%%%%%%%%%%%%%%%%%%%%%%%%%%%%%%%%%%%%%%%%%%%%%
\section{Short-time linear response and two-time correlators}
\label{sec:linear_response}

Here we discuss the appearance of two-time correlation functions in
the short-time linear response of correlators to perturbations of a
Hamiltonian.  Consider an initial Hamiltonian $H$ perturbed by an
operator $V$ with $\norm{V}\ll\norm{H}$, where $\norm{\O}$ denotes the
operator norm of $\O$, such that the net Hamiltonian is
$\tilde H=H+V$.  We denote the generator of Heisenberg time evolution
under the perturbed (unperturbed) Hamiltonian by $\tilde T$ ($T$).
These generators are related by
\begin{align}
  \tilde T = T + i\ul V
\end{align}
where $\ul V$ is a superoperator whose action on operators $\O$ is
defined by
\begin{align}
  \ul V \O \equiv \sp{V,\O}_-.
\end{align}
Through quadratic order in the time $t$ and linear order in the
perturbation $\ul V$, we can say that
\begin{align}
  e^{t\tilde T}
  \approx \f12\sp{e^{tT}, e^{it\ul V}}_+
  \approx e^{tT} + \f12 it \sp{e^{t T}, \ul V}_+.
\end{align}
Defining perturbed and unperturbed Heisenberg operators
$\tilde\O\p{t}$ and $\O\p{t}$ that respectively satisfy
$\bk{\tilde\O\p{t}}=\bk{e^{t\tilde T}\O}$ and
$\bk{\O\p{t}}=\bk{e^{tT}\O}$, we thus find that for sufficiently small
times $t$ and weak perturbations $V$,
\begin{align}
  \bk{\tilde\O\p{t} - \O\p{t}}
  = \Bk{\p{e^{t\tilde T} - e^{tT}} \O}
  \approx \f12 i t \p{\bk{\sp{V,\O}_-\p{t}}
    + \bk{\sp{V,\O\p{t}}_-}}.
  \label{eq:response}
\end{align}
Two-time correlators $\bk{V\O\p{t}}$ and $\bk{\O\p{t}V}$, in addition
to the expectation values $\bk{\p{V\O}\p{t}}$ and
$\bk{\p{\O V}\p{t}}$, thus determine the short-time linear response of
correlators $\bk{\O\p{t}}$ to perturbations $V$ of a Hamiltonian.

\bibliography{\jobname}

\end{document}